\documentclass[preprint,journal]{vgtc}

\usepackage{mathptmx}
\usepackage{graphicx}
\usepackage{times}
\usepackage{xcolor}
\usepackage{wrapfig}
\usepackage{caption}
\usepackage{subcaption}
\usepackage{soul}
\usepackage{multirow}
\usepackage{tikz}
\usepackage{enumitem}
\usepackage{rotating}
\usepackage{adjustbox}
\usepackage{standalone}

\newcommand{\tikzcircle}[2][black,fill=black]{\tikz[baseline=-0.5ex]\draw[#1,radius=#2] (0,0) circle ;}%
\setlist[description]{%
leftmargin=\parindent,
labelindent=0pt,
topsep=0.25em,
parsep=0.25em
}
\newcommand{\inlinegraphics}[1]{%
  \begingroup\normalfont
  \includegraphics[height=\fontcharht\font`\B]{#1}%
  \endgroup
}

\definecolor{col-XAI}{RGB}{22,83,127}
\definecolor{col-IML}{RGB}{192,0,0}
\newcommand{\xai}[1]{\textcolor{col-XAI}{#1}}
\newcommand{\iml}[1]{\textcolor{col-IML}{#1}}
\usepackage[bookmarks,backref=true,linkcolor=black]{hyperref}
\hypersetup{
  pdfauthor = {},
  pdftitle = {},
  pdfsubject = {},
  pdfkeywords = {},
  colorlinks=true,
  linkcolor= black,
  citecolor= black,
  pageanchor=true,
  urlcolor = black,
  plainpages = false,
  linktocpage
}

\onlineid{1076}

\vgtccategory{Research}
\ieeedoi{10.1109/TVCG.2019.2934629}

\title{\textbf{explAIner}: A Visual Analytics Framework\\ for Interactive and Explainable Machine Learning}
\author{Thilo Spinner, Udo Schlegel, Hanna Sch\"afer, and Mennatallah El-Assady}
\authorfooter{T. Spinner, U. Schlegel, H. Sch\"afer, and M. El-Assady are with the University of Konstanz. E-Mail: firstname.lastname@uni-konstanz.de}

\shortauthortitle{Spinner \MakeLowercase{\textit{et al.}}:
explAIner: A Visual Analytics Framework for Interactive and Explainable Machine Learning}

\abstract{
We propose a framework for interactive and explainable machine learning that enables users to (1)  understand machine learning models; (2) diagnose model limitations using different explainable AI methods; as well as (3) refine and optimize the models.
Our framework combines an iterative XAI pipeline with eight global monitoring and steering mechanisms, including quality monitoring, provenance tracking, model comparison, and trust building.
To operationalize the framework, we present explAIner, a visual analytics system for interactive and explainable machine learning that instantiates all phases of the suggested pipeline within the commonly used TensorBoard environment.
We performed a user-study with nine participants across different expertise levels to examine their perception of our workflow and to collect suggestions to fill the gap between our system and framework.
The evaluation confirms that our tightly integrated system leads to an informed machine learning process while disclosing opportunities for further extensions.
}
\keywords{Explainable AI, Interactive Machine Learning, Deep Learning, Visual Analytics, Interpretability, Explainability}
\teaser{
    \centering
    \includegraphics[width=0.975\textwidth]{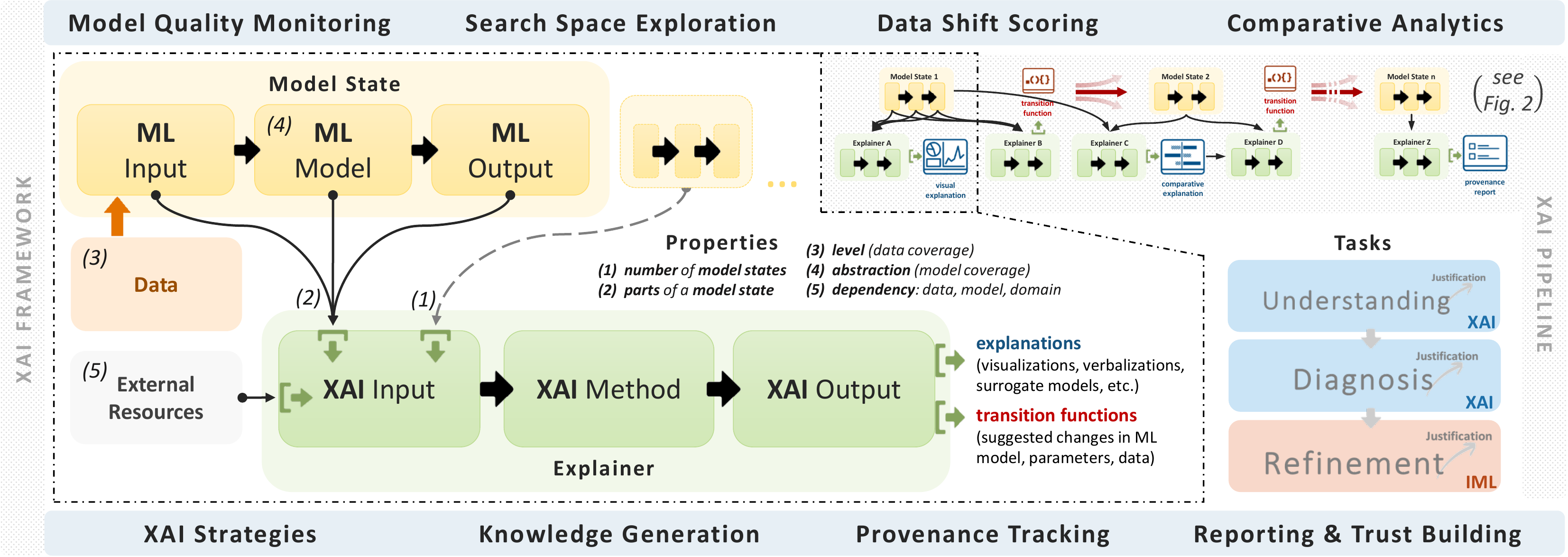}
    \caption{ 
    Close-up view of an \textit{explainer}, the main building-block  used to construct an iterative \textit{XAI pipeline} for the \textit{understanding}, \textit{diagnosis}, and \textit{refinement} of ML models.
    Explainers have five properties;
    they take one or more model states as input, applying an XAI method, to output an explanation or a transition function.
    Global monitoring and steering mechanisms expand the pipeline to the full \textit{XAI framework}, supporting the overall workflow by guiding, steering, or tracking the explainers during all steps.
    }
    \label{fig:teaser}
    \vspace{1em}
}

\begin{document}
\firstsection{Introduction}

\maketitle

Since the first presentation of neural networks in the 1940s~\cite{mcculloch1943logical}, we have seen a great increase in works on Artificial Intelligence (AI) and Machine Learning (ML). Especially within the last decade, computational resources have become cheaper and more accessible.
This development has led to new state-of-the-art solutions, e.g., Deep Learning (DL), while the increasing availability of tools and libraries has led to a democratization of ML methods in a variety of domains~\cite{Hohman2018}. For example, DL methods outperform traditional algorithms for image processing~\cite{rawat2017deep} or natural language processing~\cite{young2018recent} and can often be applied by domain experts without prior ML expertise~\cite{chollet2015keras}.

Despite the significant improvement in performance, DL models create novel challenges, due to their nature of being \textit{black-boxes}~\cite{Voosen2017}. For model developers, missing transparency in the decision-making of DL models often leads to a time-consuming trial and error process~\cite{Young2015}. Additionally, whenever such decisions concern end-user applications, e.g., self-driving cars, trust is essential. In critical domains, this trust has to be substantiated either by reliable and unbiased decision outcomes, or convincing rationalization and justifications~\cite{Keng2018}. The growing prevalence of AI in security-critical domains leads to an ever-increasing demand for explainable and reproducible results.

Several solutions address the problem of missing transparency in black-box models, often referred to as e\textbf{X}plainable \textbf{A}rtificial \textbf{I}ntelligence (XAI)~\cite{DARPA2016}. Even though AI algorithms often cannot be directly explained~\cite{Adadi2018}, XAI methods aim to provide human-readable, as well as interpretable explanations of the decisions taken by such algorithms. XAI is further driven by newly introduced regulations, such as the \textit{European General Data Protection Regulation}~\cite{GDPR2018}, demanding accessible justifications for automated, consumer-facing decisions, prompting businesses to seek reliable XAI solutions. A natural way to obtain human interpretable explanations is through visualizations.

More recent work focuses not only on visual design but also on interactive, mixed-initiative workflows, as provided by Visual Analytics (VA) systems~\cite{Endert2017}. Also, an exploratory workflow~\cite{Sacha2019} enables a more targeted analysis and design of ML models. Visual analytics further helps in bridging the gap between user knowledge and the insights XAI methods can provide. As AI is affecting a broader range of user groups, ranging from everyday users to model developers, the differing levels of background knowledge in these user groups bring along varying requirements for the explainability.

There has been extensive theoretical work on the role of visual analytics in deep learning~\cite{Hohman2018}, as well as the synergetic effects this combination can generate~\cite{Endert2017}. The fields of \textit{interactive}~\cite{Jiang2018iml}, \textit{interpretable}~\cite{Doshi-Velez2017towardsrigorous}, as well as \textit{explainable}~\cite{Adadi2018} ML are also well-studied. While these works bring up a variety of best-practices and theoretical descriptions, they often lack a tight integration into a practical framework.
In this paper, we propose a visual analytics framework for interactive and explainable ML that combines the essential aspects of previous research.
Our work is designed to target three user groups. Primarily, we focus on \emph{model users} and \emph{model developers}, as outlined by Hohman et al.~\cite{Hohman2018}. These two user groups are familiar with using and/or developing ML models and are, hence, interested in understanding, diagnosing, as well as refining such models in a given application context~\cite{Liu2017}. Our third user group, however, are \emph{model novices}. These are non-experts in ML, interested in understanding ML concepts and getting to know more about applying ML models, e.g., for specific domains. Such an educational use of our framework is facilitated through user guidance and interaction monitoring.  
End-consumers of AI products are not considered separately. Our \textit{XAI framework} is built upon an \textit{XAI pipeline} that is designed to enable the iterative process of model understanding, diagnosis, and refinement. In addition, to support these three tasks, \textit{global monitoring and steering mechanisms} (\autoref{subsec:global-monitoring-and-steering-mechanisms}) assist the overall explanation process. \autoref{fig:teaser} depicts a close-up view of an \textit{explainer}, the main building-block of the pipeline.

In recent research, a variety of concrete XAI methods and implementations have been proposed. However, these tools often are implemented as standalone prototype solutions, lacking an integration into the active ML developing and debugging process. Therefore, a large gap between theory and practice has arisen. As confirmed by our study, most people who are involved in the model usage and development process are familiar with the general concepts of  XAI, but most do not have extensive  hands-on experience using such tools.
Therefore, in contrast to previous work, we not only want to describe the theoretical workflow but use the framework to \textit{operationalize} these concepts in a system implementation, called \textit{explAIner}. 
We decided to integrate our system in TensorBoard (TB), since it is an established tool when it comes to the analysis of DL models. Our system provides an interactive exploration of the model graph, on-demand aggregation, and visualization of performance metrics as well as an integration of high-level explainers such as \textit{LIME}~\cite{Ribeiro2016lime} or \textit{LRP}~\cite{Bach2015lrp}. Based on our framework, our system follows the XAI pipeline and integrates parts of the proposed global monitoring components, such as user guidance.

Finally, we evaluate the implemented system in a qualitative user-study, with nine participants, ranging from model novices to model developers.
During the pair analytics sessions~\cite{kaastra2014field}, we analyze the usefulness of our tool while deriving ideas for future versions.

Summarizing, the contribution of this paper is threefold:
\textbf{(1)}~We propose a \textbf{conceptual framework} describing a generalizable workflow for interactive and explainable ML.
\textbf{(2)}~We present \textbf{explAIner}, a real-world system implementation, based on the proposed framework.
\textbf{(3)}~Finally, we evaluate our approach in a \textbf{user-study} with participants across different expertise levels to assess the quality of our approach and its influence on their workflow.

\begin{table}[]

\small
\centering
\begin{tabular}{|l|c|c|c|c|c|c|c|}
\hline
\multicolumn{1}{|c|}{\multirow{2}{*}{XAI method}}       & \multicolumn{2}{c|}{Level}                                                                                                                               & \multicolumn{2}{c|}{Abstraction}                     & \multicolumn{3}{c|}{Dependency}                                                       \\ 
\multicolumn{1}{|c|}{}                                                            & \multicolumn{1}{c|}{Global}                                              & \multicolumn{1}{c|}{Local}                                                          & \multicolumn{1}{c|}{Low} & \multicolumn{1}{c|}{High} & \multicolumn{1}{c|}{Data} & \multicolumn{1}{c|}{Model} & \multicolumn{1}{c|}{Domain} \\ \hline
LIME~\cite{Ribeiro2016lime} & \tikzcircle[black, fill=white]{2pt} & \tikzcircle{2pt} &\tikzcircle[black, fill=white]{2pt} & \tikzcircle{2pt} & \tikzcircle{2pt} & \tikzcircle[black, fill=white]{2pt} & \tikzcircle[black, fill=white]{2pt} \\ \hline
ANCHORS~\cite{anchors}         & \tikzcircle[black, fill=white]{2pt} & \tikzcircle{2pt} & \tikzcircle[black, fill=white]{2pt} & \tikzcircle{2pt} & \tikzcircle{2pt} & \tikzcircle[black, fill=white]{2pt} & \tikzcircle[black, fill=white]{2pt} \\ \hline
CAV.~\cite{Kim2018tcav} & \tikzcircle[black, fill=white]{2pt} & \tikzcircle{2pt} & \tikzcircle{2pt} & \tikzcircle{2pt} & \tikzcircle{2pt} & \tikzcircle{2pt} & \tikzcircle{2pt} \\ \hline
e-LRP~\cite{Bach2015lrp} & \tikzcircle[black, fill=white]{2pt} & \tikzcircle{2pt}  & \tikzcircle[black, fill=white]{2pt} & \tikzcircle{2pt} & \tikzcircle{2pt} & \tikzcircle{2pt} & \tikzcircle{2pt} \\ \hline
z-LRP~\cite{Ancona2017deeplift} & \tikzcircle[black, fill=white]{2pt} & \tikzcircle{2pt}  & \tikzcircle[black, fill=white]{2pt} & \tikzcircle{2pt} & \tikzcircle{2pt} & \tikzcircle{2pt} & \tikzcircle{2pt} \\ \hline
DeepTaylor~\cite{Montavon2017taylorlrp} & \tikzcircle[black, fill=white]{2pt} & \tikzcircle{2pt}  & \tikzcircle[black, fill=white]{2pt} & \tikzcircle{2pt} & \tikzcircle{2pt} & \tikzcircle{2pt} & \tikzcircle{2pt} \\ \hline
Saliency~\cite{Simonyan2013saliency} & \tikzcircle[black, fill=white]{2pt} & \tikzcircle{2pt}  & \tikzcircle[black, fill=white]{2pt} & \tikzcircle{2pt} & \tikzcircle{2pt} & \tikzcircle{2pt} & \tikzcircle{2pt} \\ \hline
Gradient~\cite{Alber2018} & \tikzcircle[black, fill=white]{2pt} & \tikzcircle{2pt}  & \tikzcircle[black, fill=white]{2pt} & \tikzcircle{2pt} & \tikzcircle{2pt} & \tikzcircle{2pt} & \tikzcircle{2pt} \\ \hline
DeepLIFT~\cite{Shrikumar2017deeplift} & \tikzcircle[black, fill=white]{2pt} & \tikzcircle{2pt}  & \tikzcircle[black, fill=white]{2pt} & \tikzcircle{2pt} & \tikzcircle{2pt} & \tikzcircle{2pt} & \tikzcircle{2pt} \\ \hline
grad*input~\cite{Shrikumar2016gradient*input} & \tikzcircle[black, fill=white]{2pt} & \tikzcircle{2pt}  & \tikzcircle[black, fill=white]{2pt} & \tikzcircle{2pt} & \tikzcircle{2pt} & \tikzcircle{2pt} & \tikzcircle{2pt} \\ \hline
Grad-CAM~\cite{Selvaraju2017grad-cam} & \tikzcircle[black, fill=white]{2pt} & \tikzcircle{2pt}  & \tikzcircle[black, fill=white]{2pt} & \tikzcircle{2pt} & \tikzcircle{2pt} & \tikzcircle{2pt} & \tikzcircle{2pt} \\ \hline
Occlusion~\cite{Zeiler2014deconvnet} & \tikzcircle[black, fill=white]{2pt} & \tikzcircle{2pt}  & \tikzcircle[black, fill=white]{2pt} & \tikzcircle{2pt} & \tikzcircle{2pt} & \tikzcircle{2pt} & \tikzcircle{2pt} \\ \hline
SmoothGrad~\cite{Smilkov2017smoothgrad} & \tikzcircle[black, fill=white]{2pt} & \tikzcircle{2pt}  & \tikzcircle[black, fill=white]{2pt} & \tikzcircle{2pt} & \tikzcircle{2pt} & \tikzcircle{2pt} & \tikzcircle{2pt} \\ \hline
Integrated Grad~\cite{Sundararajan2017integratedgradients} & \tikzcircle[black, fill=white]{2pt} & \tikzcircle{2pt}  & \tikzcircle[black, fill=white]{2pt} & \tikzcircle{2pt} & \tikzcircle{2pt} & \tikzcircle{2pt} & \tikzcircle{2pt} \\ \hline
DeConvNet~\cite{Zeiler2010deconvnet} & \tikzcircle[black, fill=white]{2pt} & \tikzcircle{2pt}  & \tikzcircle[black, fill=white]{2pt} & \tikzcircle{2pt} & \tikzcircle{2pt} & \tikzcircle{2pt} & \tikzcircle{2pt} \\ \hline
Node-Link Vis~\cite{Harley2015imageinterpreter} & \tikzcircle{2pt} & \tikzcircle{2pt}  & \tikzcircle{2pt} & \tikzcircle[black, fill=white]{2pt} & \tikzcircle{2pt}/\tikzcircle[black, fill=white]{2pt} & \tikzcircle{2pt} & \tikzcircle[black, fill=white]{2pt} \\ \hline
Info Flow~\cite{Wongsuphasawat2018} & \tikzcircle{2pt} & \tikzcircle[black, fill=white]{2pt}  & \tikzcircle{2pt} & \tikzcircle[black, fill=white]{2pt} & \tikzcircle{2pt}/\tikzcircle[black, fill=white]{2pt} & \tikzcircle{2pt} & \tikzcircle[black, fill=white]{2pt} \\ \hline
MinMax* & \tikzcircle[black, fill=white]{2pt} & \tikzcircle[black, fill=white]{2pt}  & \tikzcircle{2pt} & \tikzcircle[black, fill=white]{2pt} & \tikzcircle[black, fill=white]{2pt} & \tikzcircle{2pt} & \tikzcircle[black, fill=white]{2pt} \\ \hline
HistoTrend* & \tikzcircle[black, fill=white]{2pt} & \tikzcircle[black, fill=white]{2pt}  & \tikzcircle{2pt} & \tikzcircle[black, fill=white]{2pt} & \tikzcircle[black, fill=white]{2pt} & \tikzcircle{2pt} & \tikzcircle[black, fill=white]{2pt} \\ \hline
Dead Weight* & \tikzcircle{2pt} & \tikzcircle[black, fill=white]{2pt}  & \tikzcircle{2pt} & \tikzcircle[black, fill=white]{2pt} & \tikzcircle[black, fill=white]{2pt} & \tikzcircle{2pt} & \tikzcircle[black, fill=white]{2pt} \\ \hline
Saturated Weight* & \tikzcircle{2pt} & \tikzcircle[black, fill=white]{2pt}  & \tikzcircle{2pt} & \tikzcircle[black, fill=white]{2pt} & \tikzcircle[black, fill=white]{2pt} & \tikzcircle{2pt} & \tikzcircle[black, fill=white]{2pt} \\ \hline
\end{tabular}
\caption{Properties of XAI methods. \textbf{Level} is the data coverage: local (sample) or global (full dataset). \textbf{Abstraction} is the model coverage: high (full model) or low (model parts). \textbf{Dependency} specifies neccessary inputs for explainer. (*) are our own implementations.}
\label{tab:xai}
\vspace{-1em}
\end{table}

\section{Related Work}
To design XAI framework, we collected important concepts from several surveys, system, and position papers that focus either on XAI, Interactive Machine Learning (IML), or VA. We also classify a selection of relevant \textit{XAI methods} according to their properties, as shown in \autoref{tab:xai}.
These XAI methods are available as different explainers in our system implementation. 
Further, we review existing VA and IML tools and classify~\footnote{An overview of this classification is provided in Table S1 (supplementary).} them according to their applicability to the tasks of our XAI pipeline, as well as the input they are operating on. 
\subsection{Previous Conceptual Work}
To derive the concepts presented in our XAI framework, we surveyed the previous work that proposed and discussed conceptual relations.

The recent survey by Adadi and Berrada~\cite{Adadi2018} provides an entry point to XAI, covering basic concepts, existing methods, and future research opportunities.
However, they identify a lack of formalism in the field and demand ``clear, unambiguous definitions,'' revealing a research gap for a conceptual framework structuring the XAI process, which we intend to fill.
While Doshi-Velez and Kim~\cite{Doshi-Velez2017towardsrigorous} provide definitions for interpretability, they observe a need for real-world applications. The directions they give for establishing a general and multifaceted model are considered by our framework and subsequent system implementation.
By summarizing XAI motivations and characteristics, Guidotti et al.~\cite{Guidotti2018} present an extensive overview of XAI, especially current methods for the explanation of models, which our framework incorporates to tackle the latest challenges black-box models impose.

Regarding interactive machine learning, recent developments are captured by Jiang and Liu~\cite{Jiang2018iml}, who identify open research questions, including explanations for model novices, as well as global monitoring of the analytical process of explainability. These build the foundation for the monitoring and steering mechanisms of our framework.
In the context of visual analytics, Liu et al.~\cite{Liu2017} structure the IML workflow into the three tasks of understanding, diagnosis, and refinement, which we utilize to structure the process of explainability in our XAI pipeline.
In our framework, we focus on deriving synergies from this combination, e.g., by closing the ML loop between diagnosis and refinement. Such synergistic effects have been recently surveyed by Endert et al., who summarize recent advances on the integration of ML into VA~\cite{Endert2017}, such as combining interactive visualization approaches and controllable ML algorithms. 
Focusing on VA in the field of deep learning, Hohman et al.~\cite{Hohman2018} use an interrogative survey method to categorize recent work according to the six W-Questions~\cite{Hart2002}. Based on their discussion of research opportunities, we decided to target the three user groups and the four goals (interpretability, debugging, comparing, and education) they identify.

\subsection{Approaches for Explainable AI}

To support a wide range of relevant explainers, we reviewed a variety of XAI methods and classified their characteristics based on the properties highlighted in \autoref{tab:xai}. All reviewed XAI methods are supported by our framework, and most of them are part of the explAIner system.

One popular method is Local Interpretable Model-Agnostic Explanations (LIME)~\cite{Ribeiro2016lime}. It uses the models' output on a data sample to generate a linear surrogate model that explains the feature importance. A similar technique, ANCHORS~\cite{anchors}, additionally focuses the most influential input areas, so-called anchors, to formalize decision rules. Both methods do not consider the underlying model (model-agnostic) but use the sample inputs and outputs of the model (data-dependent) to explain a (local-level) decision boundary generated by the complete model (high abstraction). They can be applied domain independently.

A different type of XAI methods is represented by Saliency Maps~\cite{Simonyan2013saliency}.
They build a visual representation for feature importance by highlighting aspects in each sample as a mask of how the model perceives its input data~\cite{Guidotti2018}.
In contrast to LIME and Anchors, they are only used on artificial neural networks (ANNs) (model-specific). There are more techniques to improve the results of saliency maps, such as gradient*input~\cite{Shrikumar2016gradient*input}, SmoothGrad~\cite{Smilkov2017smoothgrad}, Integrated Gradients~\cite{Sundararajan2017integratedgradients}, Grad-CAM~\cite{Selvaraju2017grad-cam}, and DeepLIFT~\cite{Shrikumar2017deeplift}. All of these techniques use a data sample (data-dependent) from the image or text domain (domain-dependent) on ANNs (model-specific) to explain a (local-level) decision generated by the complete model (high abstraction). 
Additionally, there are two more prominent methods that have the same characterization but slightly different techniques. Layer-wise Relevance Propagation (LRP)~\cite{Bach2015lrp, Montavon2017taylorlrp} abstractly propagates a score from the output to the input to show significant features (e.g., pixelwise contribution). DeConvNet~\cite{Zeiler2010deconvnet} maps features on pixels to show the reverse activation of convolutional layers.

In contrast to these high-abstraction methods, Concept Activation Vectors (CAVs)~\cite{Kim2018tcav} operate on concrete network layers. This XAI method enables users to verify \textit{how} their (data-dependent) understanding of a concept's importance (e.g., stripes) is represented in the ANNs (model-specific) prediction (e.g. zebra) for a sample (local-level) image (domain-dependent) in each or all layers (low+high abstraction).

Other XAI methods only allow for a low-abstraction, such as visualizing convolutional filters~\cite{Harley2015imageinterpreter}, or showing the dataflow through the computational graph~\cite{Wongsuphasawat2018}. These methods are especially useful for model developers, who want to improve their models using a low-abstraction XAI method as a quality metric. Inspired by such a use-case, we implemented further low-abstraction methods for our system, including MinMax, HistoTrend, DeadWeight, and SaturatedWeight.

As shown in our review, the existing methods cover a wide range of insights and application constraints. An ideal system for explaining ML models needs to provide a collection of different XAI methods. Hence, we reviewed the first toolbox-like interfaces that aim at combining different methods into one system. iNNvestigate~\cite{Alber2018} builds an out-of-the-box approach to use saliency masks on given DNNs. A similar system is DeepExplain~\cite{Ancona2017deeplift},  which provides improved algorithms and implementations for LRP and DeepLIFT.
However, these approaches provide only some XAI methods without an interactive machine learning workflow. In our system implementation we support all explainer types simultaneously and embed them in an IML workflow.

\subsection{Interactive Machine Learning and Visual Analytics}

Our conceptual framework aims at covering not only different XAI methods but also the process of iterative and interactive explanation by combining IML and VA.
In contrast to fully automated approaches such as AutoML~\cite{Bergstra2013automl} or Neural Architecture Search~\cite{Elsken2018nnsearch}, IML strives to incorporate the human into the model building, training, and correction process to optimize a model~\cite{Fails2003}. VA can be applied to the IML workflow to boost the model development process through tailored visual interfaces~\cite{Sacha2019}. VA for IML tightly integrates the user to promote further sensemaking during the model development workflow~\cite{Endert2017}. During our review of existing IML/VA systems, we classify several solutions according to how they cover the three tasks of our pipeline~\cite{Liu2017}, as well as provenance tracking and reporting. Moreover, we show why a general XAI system, comprising all stages and tasks, is needed to address the variance in interpreting black-box models.

The \textbf{understanding} phase can be interpreted in different ways depending on the target user group.
For a model novice, some systems use VA as an educational tool to explain ML concepts. For instance, Harley~\cite{Harley2015imageinterpreter} visualizes changes of an image along with the affected layers of an ANN.  Smilkov et al.~\cite{Smilkov2016tensorflowplayground} also provide an interactive, visual representation of an ANN. Further work offers various ways to explore the graphical representation of DNNs, with Wongsuphasawat et al.~\cite{Wongsuphasawat2018} focusing on the architectural component and Kahng et al.~\cite{Kahng2018a} designing a dataflow pipeline of Generative Adversarial Networks (GANs). From these examples, we derive the need for an interactive graph visualization during the understanding phase.
In contrast to the educational goals of model novices, model users and model developers need to understand the model's inner-workings. Rauber et al.~\cite{Rauber16} focus on this aspect by visualizing the ANN training, as well as, both, neuron-neuron and neuron-data relationships.
Bilal et al.~\cite{Bilal2018} visualize the hierarchical abstraction of CNNs, highlighting the importance of multiple abstraction layers.
Representing features with low abstraction, Strobelt et al.~\cite{Strobelt2018a} explore the inner-workings of the hidden cell states, the activation, as well as ~\cite{Strobelt2018} the attention component of model structures. Based on the lessons learned from these works, we conclude that there is a need for providing tailored model explanations on different model abstractions levels.

Many VA systems address this gap by focussing on a model's \textbf{diagnosis} in an IML workflow to enable the detection of problems on different abstraction layers.
Some systems support a model-agnostic diagnosis by focusing on feature importance~\cite{Krause2016} or the reaction of the model to real~\cite{Zhang2019} or adversarial input examples~\cite{Liu}.
Others focus on specific elements, such as the neuron activation~\cite{Kahng2018}, hidden states of a cell~\cite{Ming2018} or action patterns of reinforcement learning algorithms~\cite{Wang} to allow model-specific diagnosis.
Finally, some systems visualize the dataflow~\cite{Liu2018} and decision paths~\cite{Zhao2019iforest} taken by the model to enable a model diagnosis during the training process.
While all these approaches allow for an integrated diagnosis, they fall short of addressing the identified issues in a subsequent refinement step.

Some VA systems go beyond the diagnosis phase and target the \textbf{refinement} of ML models.
We have identified works that are designed to diagnose and refine \textit{single} ML models, e.g., \cite{Liu2016,Pezzotti2017,Kwon2018}. 
Others  target \textit{multi-model} visual comparison for refinement, e.g., \cite{Murugesan2018,El-Assady2018ProgressiveFramework}.
In addition to this distinction, various interactive refinement approaches are used in iterative cycles, e.g., Cai et al.~\cite{Cai2019} on medical images or El-Assady et al.~\cite{Elassady2018ihtm,el2019semantic} for topic modeling. 
Such examples highlight the need for interactive and iterative refinement cycles in our XAI pipeline. Further, the comparative explanation and refinement of multiple models is essential for assessing the quality of different models and selecting the most suitable for a given context.

Besides, the \textbf{reporting} of results and the tracking of changes are essential elements of IML~\cite{simard2017machine}.
Three of the surveyed VA systems support these tasks. Krause et al.~\cite{Krause2018} and Ming et al.~\cite{Ming2019}, both, provide a visual representation of a feature's importance to the model output, while Sevastjanova et al.~\cite{Sevastjanova2018} support tracking the full workflow, as a mixed-initiative, active learning system. 
These approaches show that an XAI framework should go beyond these three IML tasks and incorporate global monitoring and steering mechanisms.

While all the reviewed approaches are highly-specialized to their use-cases and cover the respective phases of the IML workflow, we propose a pipeline that can cover different pathways through all of the addressed tasks. To aid this pipeline, global monitoring and steering mechanisms can support and guide the overall process of IML. 

\begin{figure*}
    \centering
    \includegraphics[width=\textwidth]{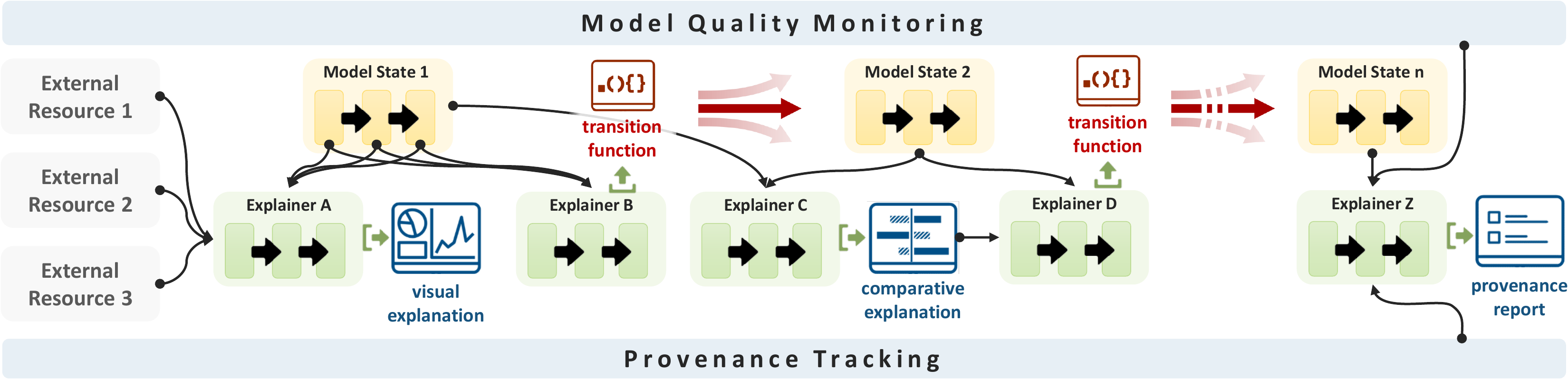}
    \caption{Unrolled view of an iterative model 
    explanation and refinement workflow. Different explainers are applied to a \textit{model state}~$1$, to generate insights and identify flaws. From the detected issues, refinements are proposed, which induce a \textit{transition} to a new \textit{model state}~$2$, etc. 
    Global monitoring and steering mechanisms are used to document the \textit{provenance} of the XAI process and keep track of different \textit{quality metrics}.}
    \label{fig:workflow}
    \vspace{-10pt}
\end{figure*}

\section{Conceptual Framework}\label{sec:conceptual-framework}

Accompanying the abundance of machine learning methods came attempts for the formalization of interactive and explainable machine learning~\cite{Endert2017}. Most of them were driven by theoretical deductions, i.e., based on surveying the literature to derive a conceptual model~\cite{Liu2017}.
However, to support the development of an interactive and explainable machine learning system, we require a conceptual model that takes the implementation needs into account, while being compatible with the proposed theoretical models.
Hence, in this paper, we propose a conceptual framework that is tailored to advance the \textit{operationalization} of interactive and explainable machine learning. Our framework is not limited by specific software or hardware constraints, but  primarily focuses on practicability, completeness, as well as full coverage.

As depicted in \autoref{fig:teaser}, an XAI pipeline constitutes the heart of our framework. This pipeline, an unrolled view of the iterative model development and optimization process, is designed to enable the understanding, diagnosis, and refinement of machine learning models \cite{Liu2017} using, so-called, \textit{explainers}. These explainer modules interact with the machine learning model to derive (1)~\textit{explanations} in the form of visualizations, verbalization, or surrogate models, as well as (2)~\textit{transition functions} for model refinement.
Enveloping the XAI pipeline are global instruments for tracking the quality and development of the explanation process, as well as enabling user guidance, provenance tracking, reporting, etc. 
In this section, we will describe our conceptual framework in more detail, starting with the XAI pipeline, followed by the \textit{global monitoring and steering mechanisms.}

\subsection{XAI Pipeline}
As depicted in \autoref{fig:workflow}, our proposed workflow constitutes multiple \textit{model states} and multiple explainers. A model state is one configuration of a trained ML model with a given set of parameters, weights, etc. Changing such parameters or weights transitions the model to a different state. 
Thus, to find an `\textit{optimal}' model for a given data and
task, we consider the search space that spans all possible model states, i.e., all different configurations of a model. Note, that changing the model architecture would result in another model with its own search space, while retraining the same model only transitions the model state. 
Using this notion, we can argue that the goal of interactive machine learning is to enable model refinement such that we transition 
 \begin{wrapfigure}[5]{l}{0.34\columnwidth}
 \vspace{-22.5pt}
  \begin{center}
    \includegraphics[width=0.34\columnwidth]{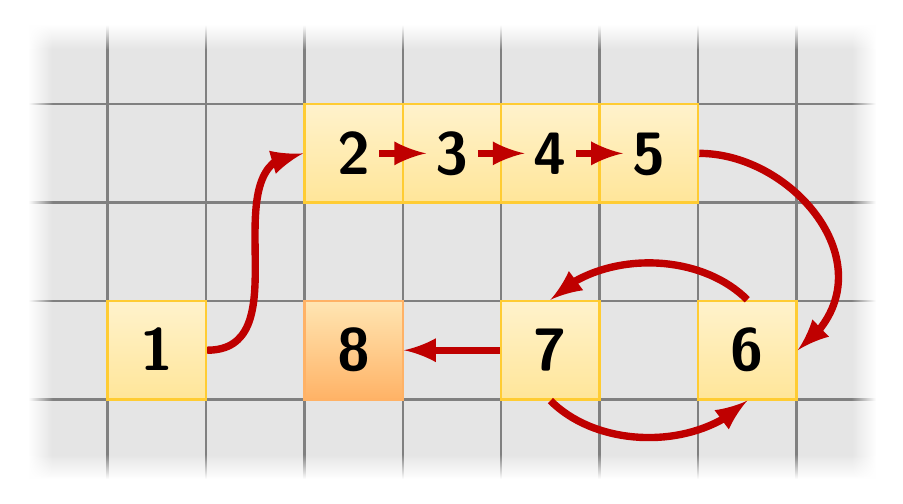}
  \end{center}
\end{wrapfigure}
to model states that are more suitable for a given problem characteristic.
More formally, we can define all operations that change a property of a model state as a transition function $f: MS_x \rightarrow MS_y $, with $MS_x$ and $MS_y$ describing two model states (which could be equivalent) within the search space. Hence, a refinement process can be seen as a traversal through multiple model states.  
Our proposed explainers are components that take into account inputs from a single model state (henceforth referred to as \textbf{single-model explainer}) or from multiple model states (henceforth referred to as \textbf{multi-model explainer}). 

Beside the \textbf{number of model states considered}, we categorize
\begin{wrapfigure}[9]{r}{0.32\columnwidth}
 \vspace{-23pt}
  \begin{center}
  \includegraphics[width=0.32\columnwidth]{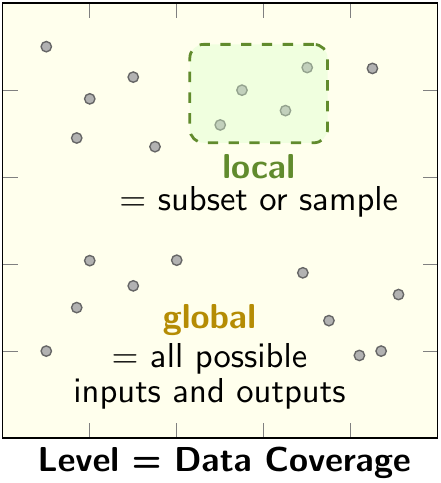}
  \end{center}
\end{wrapfigure}
explainers based on the \textbf{parts of the model state considered} as input (ML input, ML model, and/or ML output). Furthermore, we define the \textbf{explainer level} as \textit{global}, when all possible data inputs and outputs are considered. On the other hand, a \textit{local} level refers to the explainer only considering a subset or sample of the data, e.g., for explaining decision boundaries.

Additionally, we define the \textbf{explainer abstraction} as the model
 \begin{wrapfigure}[8]{l}{0.43\columnwidth}
  \vspace{-22pt}
  \begin{center}
    \includegraphics[width=0.44\columnwidth]{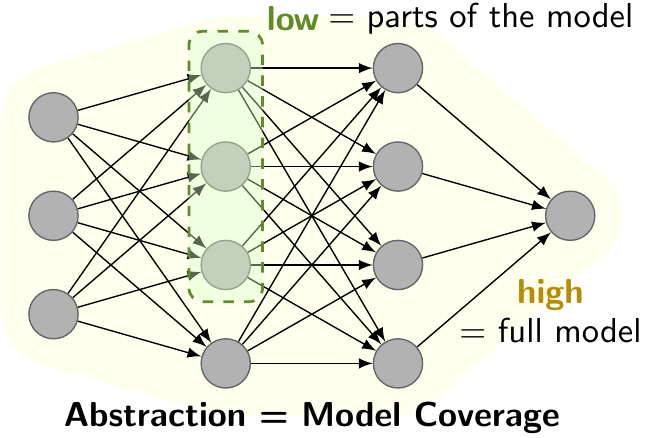}
  \end{center}
\end{wrapfigure}
coverage, i.e., a \textit{low} abstraction only considers a part of the model, while a \textit{high} abstraction considers the whole model. Lastly, each explainer can have \textbf{dependencies to the data, model, and/or domain knowledge}, i.e., resulting in explanations that are dependent on some of these factors. Generally, the output of explainers are either \xai{(1)~\textbf{explanations}} -- (interactive) visualizations, verbalizations, model surrogates, etc; or \iml{(2)~\textbf{transition functions}} 
to a new model state. 
Hence to achieve the goals of \xai{explainable (XAI)} and \iml{interactive (IML)} machine learning, \xai{explanations} are used to \xai{understand} and \xai{diagnose (XAI)} a model, and \iml{transition functions} are used for model \iml{refinement (IML)}.

\noindent
In the following we will discuss different explainer types: 

\textbf{(A)~Single-Model explainer -- } 
The most straight-forward type of
\textit{explainers} are the ones that consider inputs and outputs of a 
\begin{wrapfigure}[10]{l}{0.50\columnwidth}
 \vspace{-20pt}
  \begin{center}
    \includegraphics[width=0.51\columnwidth]{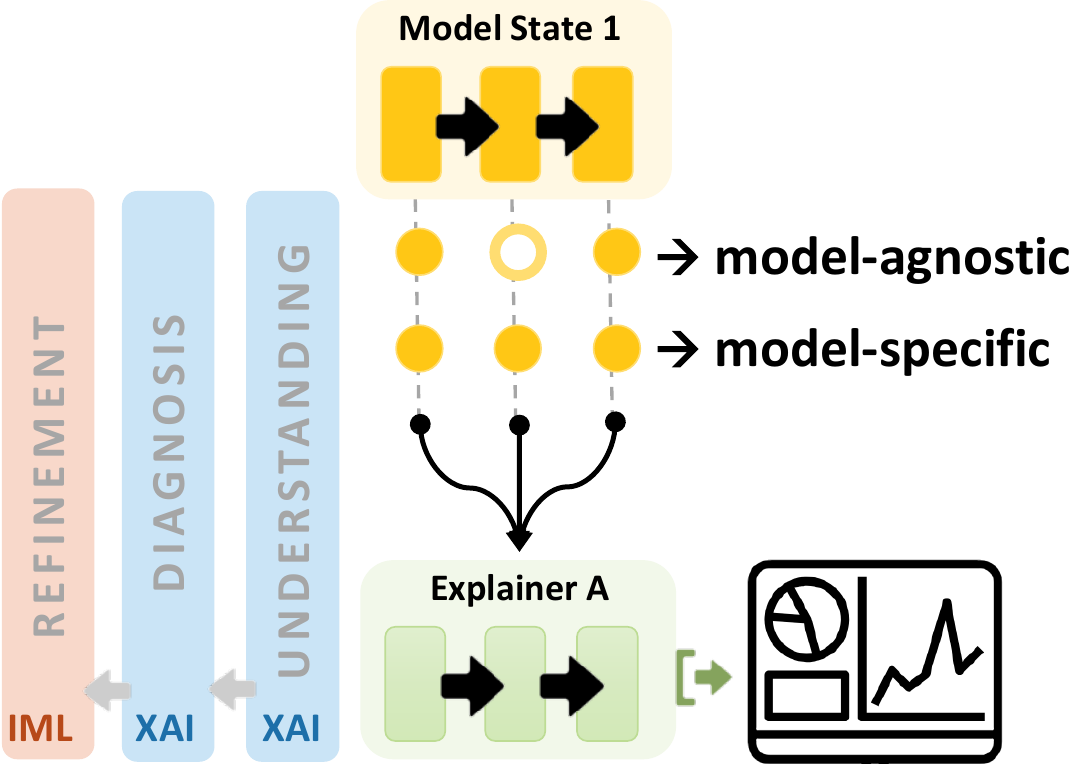}
  \end{center}
\end{wrapfigure}
machine learning model, as well as its inner-workings. These \textbf{model-specific} explainers are particularly useful for model developers as they can help in diagnosing the internal structure of a model, refining it based on the interplay of inputs and outputs, concerning the given architecture. Examples include: \cite{Elassady2018ihtm,Bach2015lrp}.
In contrast, \textbf{model-agnostic} explainers operate on the data level. They consider the model to be a black-box that performs a transition from input to output, and, thus, attempt to explain or approximate this transition. These explainers are particularly useful for model novices and model users in machine learning who are not interested in understanding the underlying model but rather in applying it to their data and tasks. Examples of such explainers include: \cite{El-Assady2018ProgressiveFramework,Ribeiro2016lime,anchors}.

Our proposed XAI pipeline can be used in different stages of the \textit{training--testing continuum} of a model by stopping or extracting a state from the current training process. However, for different user groups, some states and explainers might be more favorable. In addition to the two types mentioned above, we particularly would like to highlight explainers used for understanding the ML input, model, and output.  
Understanding the model without 
\begin{wrapfigure}[10]{r}{0.36\columnwidth}
  \vspace{-20pt}
  \begin{center}
    \includegraphics[width=0.36\columnwidth]{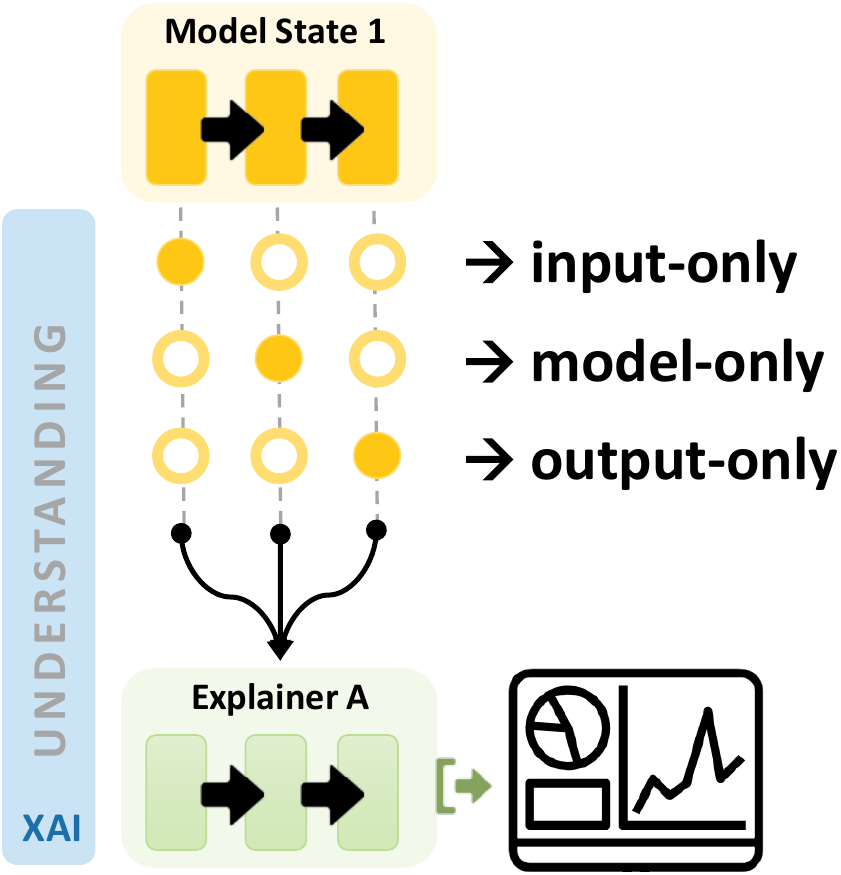}
  \end{center}
\end{wrapfigure}
considering the input and output of the data is a useful task for inspecting the model architecture and weights, as well as for educational purposes. Similarly, inspecting the input or output data distribution and characteristics is a common task. Hence, these types of explainers are particularly useful for understanding but not as much for diagnosis or refinement. One example of such a \textbf{model-only} explainer from our system implementation is the \textit{look-up explainer}, showing wiki-style entries to enable the understanding of parts of NNs. Other examples of such an explainer include: \cite{Wongsuphasawat2018,Harley2015imageinterpreter}.

\textbf{(B)~Multi-Model explainer -- } Complementary to single model 
\begin{wrapfigure}[5]{l}{0.28\columnwidth}
 \vspace{-24pt}
  \begin{center}
    \includegraphics[width=0.31\columnwidth]{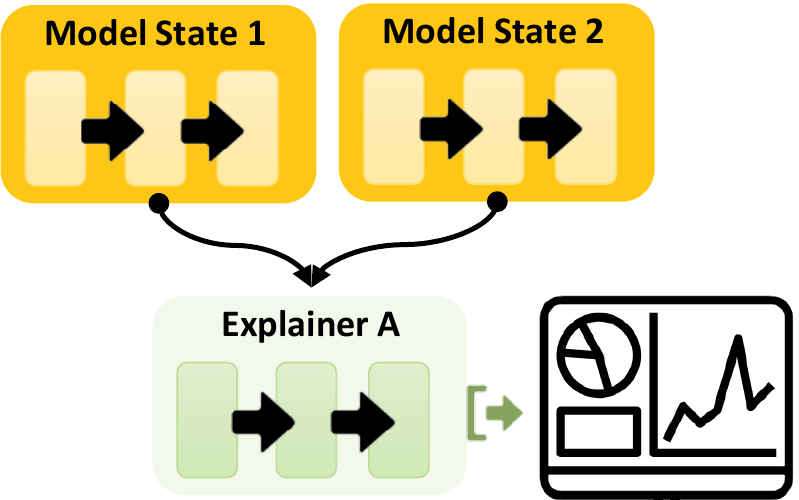}
  \end{center}
\end{wrapfigure}
explainers, multi-model ones are primarily used for the \textit{comparative}
\textit{analysis} of model states. These take as an input two (or more) model states to compare, and their output is tailored to such a task, i.e., comparative visualizations, model selection components, or a transition function based on all input states. Examples include: \cite{El-Assady2018ProgressiveFramework,Murugesan2018,Zhang2019}.

Multi-model explainer help in exploring the model state search
\begin{wrapfigure}[9]{l}{0.44\columnwidth}
 \vspace{-19pt}
  \begin{center}
    \includegraphics[width=0.42\columnwidth]{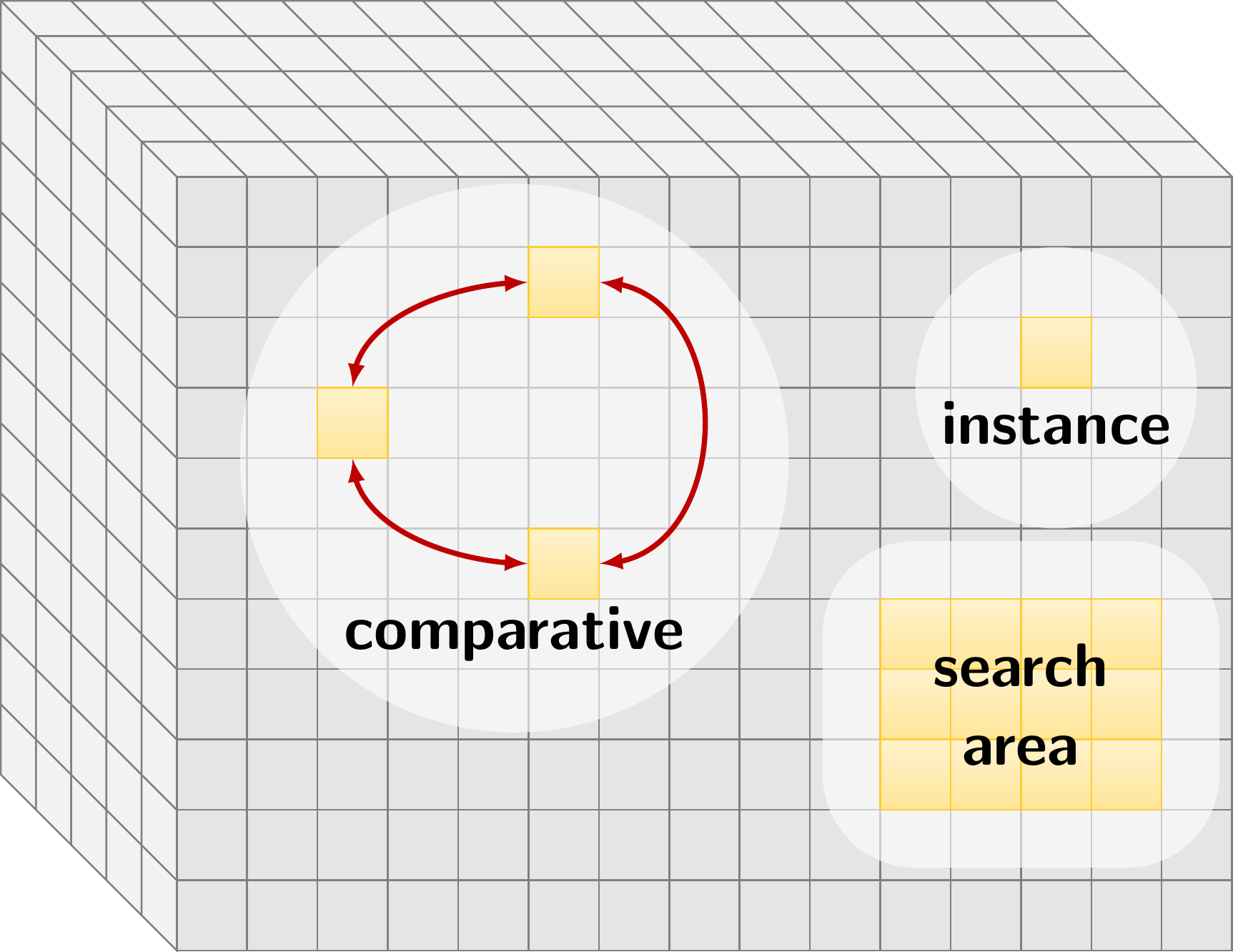}
  \end{center}
\end{wrapfigure}
space beyond single instances or points in the space. The comparative analysis, thus, enables a \textbf{point-wise inspection} of similarities and differences between a selection of different model states. In addition, we can subdivide the search space through defining a \textbf{search area} in which model states are compared.

In addition to the described explainer types, others 
might consider a different combination of inputs, in particular including \textit{model-external resources}. As an example, \autoref{fig:workflow} shows a pipeline of five explainers and three model states. The process starts with explainer $A$ that takes into account its input from three external resources, as well as model state $1$, it outputs a visual explanation which can be used for model understanding or diagnosis. On the other hand, explainer $B$ is used for refinement, as it suggests a transition function based on the same model state. The pipeline continues with explainer $C$, which is the first multi-model explainer. It takes the models state $1$ and the newly generated model state $2$ as inputs to compare.
Such a comparative analysis can enable a model selection task, or, as is the case with explainer $D$, inform a new model state. Lastly, this final model state $n$ is used for generating a provenance report based on explainer $E$. This report considers, in addition to a model state, the results of the continuous \textit{model quality monitoring} and \textit{performance tracking} for the report generation.

The proposed pipeline is subject to an adaptation to the targeted tasks and user groups. 
Our three user groups have different needs and workflows in our pipeline. (1)~ \textbf{Model novices} would see the approach as an educational tool and mostly utilize the loop between understanding and diagnosis to learn about the effects of the architectural components. (2)~\textbf{Model users} would primarily utilize the diagnosis and reporting capabilities to track, justify, and verify their decisions and interactions, while exploring the model. (3)~Lastly,  \textbf{model developers} focus on the loop between diagnosis and refinement. They are utilizing the explainers as an additional quality indicator during their IML.

\subsection{Global Monitoring and Steering Mechanisms}\label{subsec:global-monitoring-and-steering-mechanisms}
To guide, steer, and track the XAI pipeline, we propose several global mechanisms, as observed in the related work. These mechanisms ensure that users are supported in their goals during all phases of the pipeline.
This section describes the most important mechanisms, categorized into eight groups. Further aspects can be considered to tailor our framework to specific application requirements. 

\begin{description}
\itemsep -0.3pt
    \item[Model Quality Monitoring --]
      Internal performance metrics, such as accuracy, precision and recall~\cite{Powers2007metrics}, as well as measures of bias~\cite{Dietterich1995} and uncertainty~\cite{Solomatine2009} can  support the XAI pipeline in pointing the users to potential model improvement possibilities. We propose including a global monitoring component to constantly track and asses the internal quality of the different model states.

    \item[Data Shift Scoring --]
      Analogous with the continuous monitoring of the model quality, if the XAI pipeline is targeting the analysis of changing data sources, we propose the monitoring of potential data shifts~\cite{Quionero-Candela:datashiftscoring} along the \textit{training--testing continuum}. Especially for deciding when to stop the training and retraining~\cite{geman1992dilemma}, a measurement for the data fitness of the model is of immense importance.

    \item[Search Space Exploration --]
      For a targeted refinement and optimization, efficient navigation of both the model input and output spaces is vital. Examples to achieve minimum feedback for maximum gain include, for example, Speculative Execution~\cite{SBS+18}, where different potential optimizations are performed and presented to the user \textit{before} applying them on the model. 
    
    \item[Comparative Analytics --]
      Another important task is comparing and selecting models. Explainers can be designed to compare different model states on multiple levels. This, in turn, facilitates tasks, such as model selection~\cite{Fails2003}, automated model recommendation~\cite{Malkomes2016}, or automated model configuration search (AutoML)~\cite{Bergstra2013automl, Elsken2018nnsearch}, i.e., transitions from one model state to another. 
    
    \item[XAI Strategies --]
      For an adequate explanation, we propose considering global \textit{XAI Strategies}~\cite{strategies2019}. These can implement user guidance~\cite{Amershi2019} or propose the use of  alternative mediums of explanation, such as verbalization~\cite{SBE+18a}. XAI Strategies~\cite{strategies2019} structure the process of explanation through putting it into a larger context. Explainers are regarded as building blocks that use a certain explanation strategy and medium to explain an aspect. Building blocks can be connected through \textit{linear} or \textit{iterative} \textit{pathways}.
      Several explanation blocks can be grouped into a phase that is followed by a verification block to check the user's understanding of an explanation. Deciding on an effective strategy for each user group is essential. This entails configuring the amount of user guidance needed.
      
    \item[Provenance Tracking --]
      Model refinement and optimization is a ``\textit{garden of forking paths}''~\cite{gelman2013garden}. To track the temporal evolution of the user's workflow, we propose a provenance tracking component. An interaction tree~\cite{AvocadoPaper}, for example, could reveal the sequence of decisions users undertook in the XAI pipeline.

    \item[Reporting \& Trust Building --]
      To enable a reasoned justification of the user's decision-making, as well as allow for communicating the results of a workflow, we propose the implementation of reporting components. These can be used in educational settings, for example through designing them as storytelling~\cite{Chen2018} components. Such mechanisms facilitate the calibration of trust between the users and the machine learning model~\cite{MinionPaper}.

    \item[Knowledge Generation --]
      Lastly, the ultimate goal of such a visual analytics framework is knowledge generation~\cite{sacha14knowledge}. This can go in two directions, users can learn something about the ML or validate their knowledge; while models, can be taught by users~\cite{simard2017machine}, e.g., through learning from their interactions~\cite{Amershi2019}.
\end{description}

\subsection{Use-Cases}
We present one use-case for each user type to make the framework and its application more concrete. Each user is described when solving a typical task and will thus focus on specific elements of our framework.

\paragraph{Case 1}
A computer-science freshman (model novice) takes a lecture on machine learning.
As an assignment, the professor provides a neural network model which the students should explore concerning its architecture and functionality.
The framework supports the student during the entire task. 
For example, a single-model explainer, which supplies model-only explanations, could provide information about the model's architecture in the understanding task.
Using provenance tracking, the student can document his process of exploration and summarize it later in the reporting step.

\paragraph{Case 2}
Biologists (model users) want to track the movement of various animals in a zoo.
They have to choose between different off-the-shelf models to identify the animals in the images taken by cameras.
The proposed framework helps the biologists to decide, which of the models solve the task the most accurate and reliable.
Based on a labeled test dataset, the model can be diagnosed using different explainers. Thus possible explainers could be single-model explainers, which deliver model-agnostic explanations to solve the task of verification. Findings then can be directly summarized and reported to the director, justifying the decision for a specific model.

\paragraph{Case 3}
A researcher in the field of self-driving cars (model developer) has built a model which reaches an accuracy of over $99\%$ but always fails in specific situations.
The proposed framework supports the researcher in each step of the iterative model development and optimization process.
By applying different explainers to his model, he finds that his model always fails when birds are visible in the sky.
During refinement, the proposed framework proposes multiple options for state transitions by varying architecture and parameters of the model.
Using comparative analytics, the researcher can compare multiple model states based on quality metrics and applied explainers.
By iterating diagnosis and refinement, the researcher reaches an accuracy of $99.99\%$, while at the same time, he can build trust that the model is focussing on relevant parts of the cars surrounding.
Since the user wants to advance research in his field, he decides to export and share a report of his model building and explanation process.

\section{System Design and Implementation}\label{sec:system-implementation}

\begin{figure}[b]
  \includegraphics[width=\linewidth]{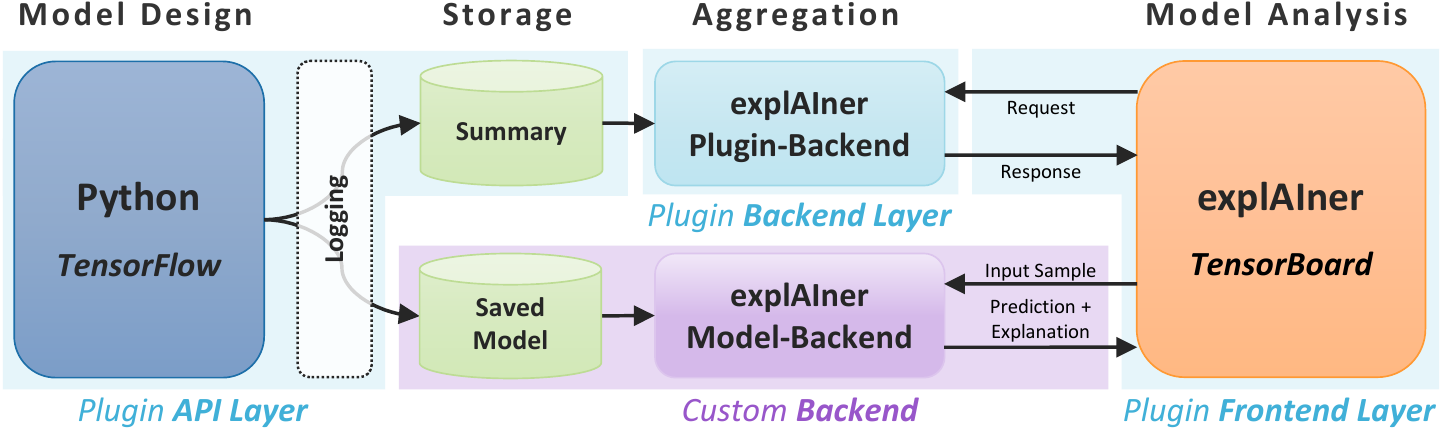}
  \caption{System design overview. The TensorFlow model is created in Python. During training, logfiles are written using the explAIner summary method, which saves graph definition, tensor contents, and the model itself. The explAIner TensorBoard plugin queries data either from the native backend implementation or from an external server, depending on whether the explainer uses tensor data or the model.}
  \label{fig:system-description-overview}
\end{figure}

Though there already exist systems including some parts of the proposed conceptual framework in \autoref{sec:conceptual-framework}, we operationalize the framework as a TensorBoard plugin called \emph{explAIner}\footnotemark.
The plugin implementation can be seen as an instantiation of the conceptual framework, translating the theoretical concepts to an actual application. 
We chose TensorBoard as the platform because it is widely used in the ML community, and our system perfectly complements and extends the native functionality it provides. More specifically, we add graph views to augment the graph entities with additional information and allow them to contain actual values and time series, which can be interactively accessed by selecting the nodes. Furthermore, we introduce a global provenance tracking component which allows to store and compare model states persistently. Finally, our system allows the execution of different XAI methods at run-time.
By design, XAI methods target specific application domains, data types, or network architectures. We address this heterogeneity by embedding explainers in the proposed VA system, which allows us to react to such constraints dynamically based on the user's needs, e.g., by showing only relevant information (filter) or proposing distinct methods over others (user guidance).

\begin{figure}[tb]
  \includegraphics[width=\linewidth]{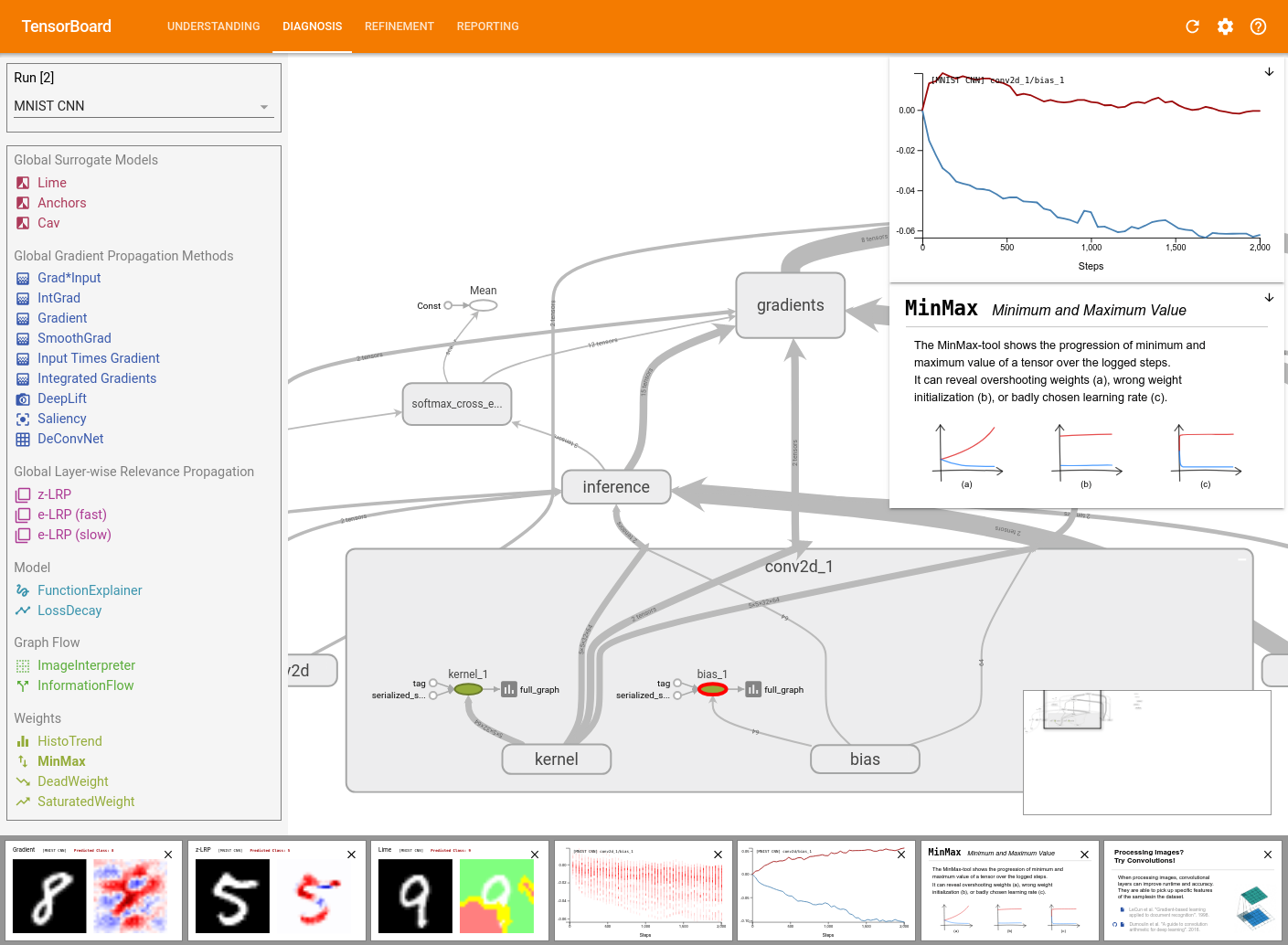}
  \caption{The \emph{diagnosis} dashboard. Explainers are arranged in a toolbox-like interface, ordered descending, from high-abstraction to low-abstraction. The graph visualization provides an overview of the full model and allows for node selection. Explanations are shown in the upper toolcard, while information about the explainer is displayed beneath. The provenance bar contains cards from interesting findings.}
  \label{fig:diagnosis-screenshot}
  \vspace{-5pt}
\end{figure}

Design decisions for our implementation are primarily guided by the proposed theoretical framework as well as TensorBoards best-practices and capabilities.
By splitting the stages of the XAI pipeline into distinct TensorBoard plugins, we aim to ensure separation of concerns~\cite{Dijkstra1982}.
Regarding UI elements, TensorBoard gave us an excellent starting point by providing reusable color scales and web components.
Furthermore, we try to stick with TensorBoards design habits, such as showing visualizations in overlaying cards (\autoref{fig:diagnosis-example}), maintaining the toolbar layout (\autoref{fig:diagnosis-screenshot}), or keeping things contained in specific tabs.
Provenance tracking is an exception: TensorBoard is not designed to have components and data shared over multiple plugin tabs, so we have to add this functionality to the TensorBoard system manually.

\autoref{fig:system-description-overview} shows an overview of the system and its components. The design and training of the TensorFlow model are done in Python manually or by an AutoML or network architecture search approach. We provide an additional explAIner summary, which can save graph definitions, tensors, and the model itself.
We store values for each tensor in the graph. 
Since the required aggregations for our explainers are known beforehand, we can transfer the aggregation step directly to the time of logging.
The size of stored data then is comparable to the summaries that are typically written using TensorFlow.
Therefore, explAIner has no significant impact on TensorBoards performance.

The explAIner plugin makes use of two different backends, depending on the explanation method.
Explainers which work on a model's inputs and outputs use an external model-backend, while explanations for graph tensors use the native TensorBoard plugin backend.

The TensorBoard developers provide an example plugin~\cite{TensorBoardPluginExample2019} as a reference for custom plugins.
A TensorBoard plugin consists of three parts which can be seen as a pipeline:
The \textbf{API layer}
defines operations to log data during model execution.
It corresponds to \emph{model design} and \emph{logging} in \autoref{fig:system-description-overview}.
The \textbf{backend layer}
loads, preprocesses, and serves the stored data.
In \autoref{fig:system-description-overview} it handles loading from \emph{storage} and \emph{aggregation}.
The \textbf{frontend layer}
queries data from the backend and renders visualizations in the UI. In \autoref{fig:system-description-overview} this is depicted as \emph{request/result} and \emph{model analysis}.
\footnotetext{System is publicly available under: \textit{http://explainer.ai/}}
These three layers have to be implemented to create a custom plugin. Using the logging operations in the API layer, we extract all relevant data from the computational graph; storage is handled by TensorFlows summary mechanism. Since TensorFlow does not provide a way to save a model as a summary, we complement the API operations by saving the model manually.
To execute the model with data, we have to bypass the automated TensorBoard backend layer.
In the frontend layer, we can query both backends with similar calls. The plugin can be injected into the TensorBoard UI by providing a custom HTML-page, which, besides the default plugins, loads our custom plugins.

We extend TensorBoard by four additional dashboard-views, one for each step of the XAI pipeline as well as one for global monitoring (reporting). The interface and interaction possibilities for each view follow the specific tasks:
\begin{description}
\itemsep -1.5pt
    \item[Understanding] provides a graph view, enabling interaction with the model to get educational information about its architecture.
    \item[Diagnosis] builds around an instanced graph view of the model, where variable nodes contain a history of their data at the logging points.
    \item[Refinement] shows interaction recommendations based on model architecture, findings from previous stages, and general heuristics.
    \item[Reporting] provides a summary of the full model analysis process. Notes on results can be arranged, annotated, and exported.
\end{description}
To keep track of the knowledge and insights generated during the complete explanation process, our system extends TensorBoard with an additional \emph{provenance bar}. It acts as a persistent clipboard and notetaking-environment, in which the user can document discoveries, thoughts, and interpretations as small provenance cards.

\subsection{Understanding}
The understanding phase is the entry point into our proposed workflow. 
For a model developer, this step offers information necessary to create a fitting model,  e.g., layer sizes, loss function, used optimizer, etc.
For a model user and a model novice, it explains a given model and its functionality by providing visual representations, descriptions and external information on the network.
In our prototype system, we implement this phase as the integration of information cards, that can be displayed by interactively focusing parts of a graph representation of the model. 
While other layouts were considered~\cite{Bischof1992,LeNail2019}, our graph view is derived from the TensorBoard graph, since it is well known in the community and reproduces TensorFlows computational graph accurately.
When hovering a graph node, a short description and explaining graphics are displayed.
Clicking on a node opens an overlay, which contains more detailed information and external references, similar to a short wiki article.
The content is manually extracted and visually appealingly prepared from wikis, blogs, and scientific publications.
Supplementary information can be retrieved for entities of different levels, ranging from the full model down to single operations.

\subsection{Diagnosis}
In the framework, we define the diagnosis phase as the most critical part of our workflow.
It enables model novices to visually explore and thus learn about the output of the model.
It offers a decision support tool for model developers, that helps them choose necessary refinements, and it gives visual feedback for verification by a model user or domain expert.
In our prototype, we offer various explainers sorted by scope, which can be interactively placed on the visual graph representation of the model.
We display the visual feedback of the explainers as overlaying cards, which can be saved to the provenance bar to trace the process of exploration.
In addition to the explainers output, we provide a second card with supplementary information on the functionality of the explainer and how its outputs can be interpreted. 
\autoref{fig:diagnosis-screenshot} shows a screenshot of the diagnosis dashboard view.

High-abstraction explainers take the trained model and a user-selected sample as input, for which they return prediction and explanation.
Low-abstraction explainers work on parts of the model and can be applied to single graph entities or a particular subset of the graph. The explanations range from time-dependent metrics of a single variable up to the visualizations of the flow of a tensor as it traverses the graph. See \autoref{tab:xai} for an overview of the explainers we implemented.
Since for more advanced networks the graph representation can become quite complex, we provide user guidance to help the user focus on interesting graph entities. This is done by visually emphasizing nodes on which a certain explainer can be applied or by marking nodes that deviate significantly from other nodes of the same type.

\begin{figure}[tb]
    \centering
    \begin{subfigure}{0.47\linewidth}
        \includegraphics[width=\linewidth]{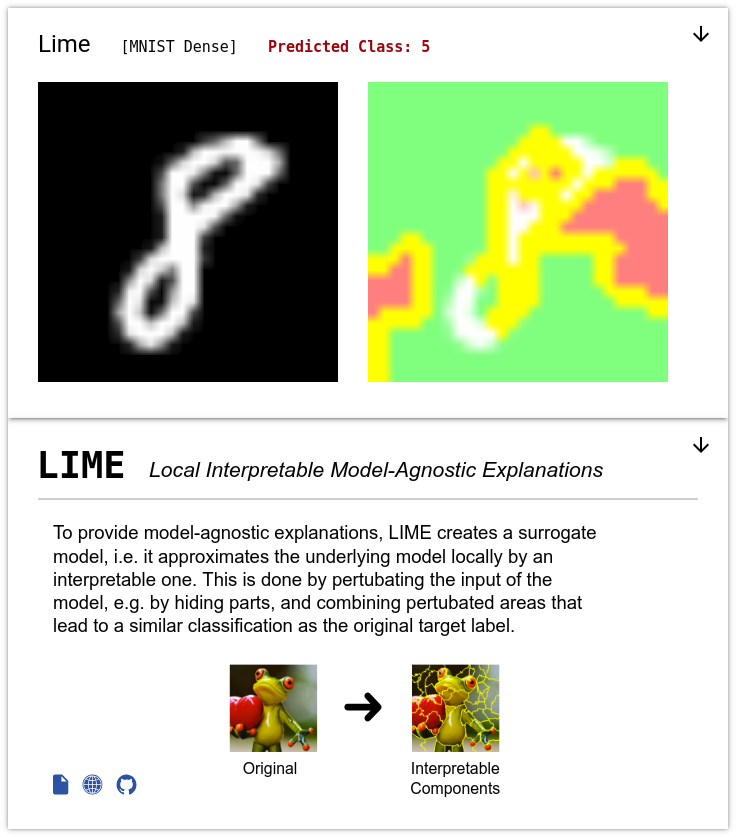}
        \caption{LIME (high-abstraction explainer)}
        \label{fig:diagnosis-example-1}
    \end{subfigure}
    \quad
    \begin{subfigure}{0.47\linewidth}
        \includegraphics[width=\linewidth]{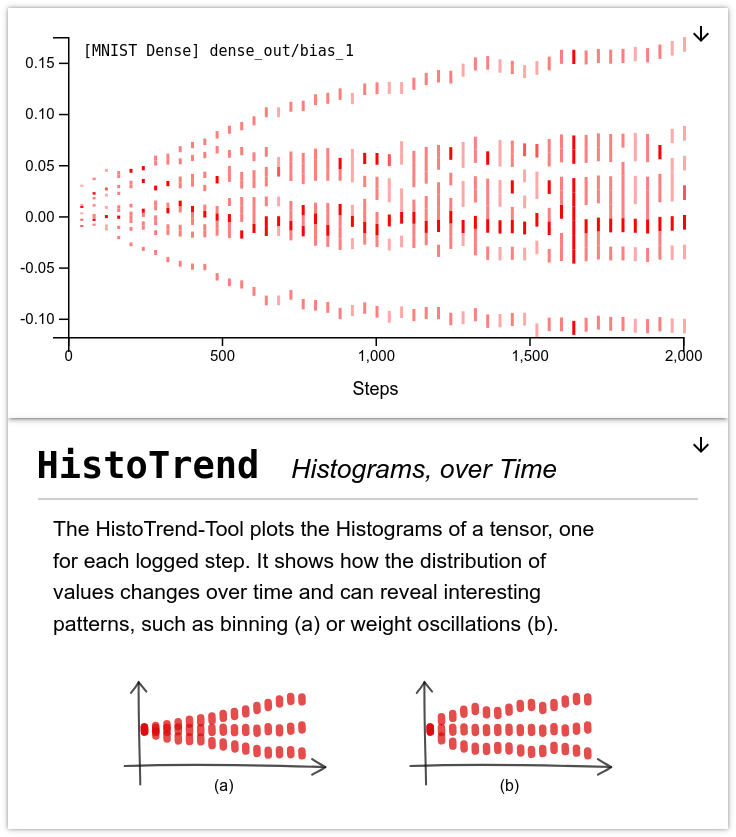}
        \caption{HistoTrend (low-abstraction explainer)}
        \label{fig:diagnosis-example-2}
    \end{subfigure}
    \caption{Information cards showing results for different explainers (top) with the corresponding descriptions for the explainer itself (bottom).}
    \label{fig:diagnosis-example}
  \vspace{-5pt}
\end{figure}

\textbf{Example For a High-Abstraction Explanation -- } When a user issues a high-abstraction explanation method, e.g., LIME~\cite{Ribeiro2016lime}, the user is prompted to select a data sample, which is sent to the model backend as input for the explanation.
The backend loads the trained model from a saved file and executes the explanation method for the given input.
After execution has completed, explanation and prediction are sent back to the explAIner frontend, where they are presented to the user (\autoref{fig:diagnosis-example-1}).
Besides LIME as an example for surrogate models, we implemented several other model-based explainers, including \emph{Layer-Wise Relevance Propagation} (LRP) and several gradient-based methods, such as \emph{Saliency} and \emph{Deeplift}~\cite{Bach2015lrp, Simonyan2013saliency, Shrikumar2017deeplift}.

\textbf{Example For a Low-Abstraction Explanation -- } Low-abstraction explanation methods can be directly applied to individual nodes of a graph.
After selecting such node, explAIner creates a request containing identifiers for node and explainer and sends it to the backend layer of the TensorBoard plugin.
The backend responds with the aggregated tensor data for the selected node and explanation.
Visualization of this data happens directly in the frontend layer of the plugin (\autoref{fig:diagnosis-example-2}).

\subsection{Refinement}

\begin{figure*}[htb]
    \centering
    \includegraphics[width=\textwidth]{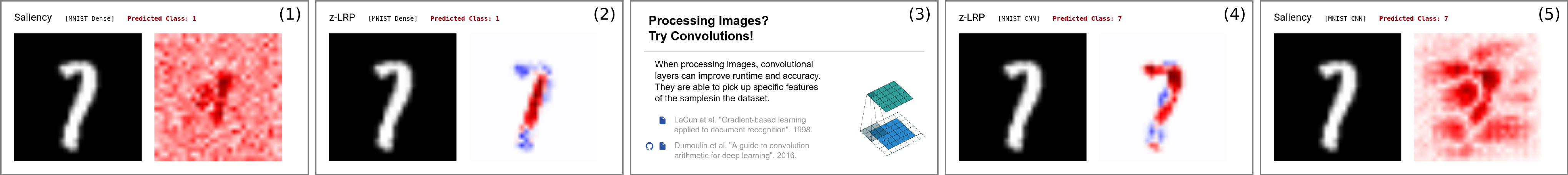}
  \caption{Use-case showing the provenance bar sequence used for model refinement towards the correct prediction of a seven. A misclassified image is analyzed using (1)~Saliency and (2)~LRP. As a refinement, explAIner suggests using (3)~convolutional layers. After applying the refinement, (4)~LRP and (5)~Saliency show a correct focus on the relevant features.}
  \label{fig:provenance-use-case}
  \vspace{-5pt}
\end{figure*}

The refinement phase is crucial for model developers that want to improve their model interactively.
For model novices and model users, this steps is more rarely utilized with the goal of further exploration.
In our prototype, we implement two different transitions into the refinement phase.
First, the user can choose to add a refinement directly related to a given explainer output.
Second, the user can directly enter the refinement phase by choosing the respective tab and choose between all possible refinements.
This transition is essential to the model developers since they might already know of more general refinements, that are dependent on the context that is given by the explainer.
Besides refinements that are targeting improvements of the model accuracy, we also focus on enhancements in space and time requirements of the model.
In this prototype, all optimization steps are supported by general textual information to help all users understand their effect.
The refinements are realized as recommendations that the model developer might follow to improve its model.
Such recommendations give a summary of how the improvement works and why the explAIner system suggests it.
Furthermore, improvements that affect the models basic functioning and therefore might change the way a model solves a specific task are provided with links to external resources.
This is meant to keep the developer up to date with the latest discoveries in AI since the field develops rapidly.
The recommendations that are suggested during the refinement step are based on heuristics, considering graph architecture, the task, that the user seems to be trying to solve, and findings from previous steps.

\subsection{Provenance Tracking and Reporting}
Our TensorBoard implementation is complemented by a \emph{provenance bar}. During the complete exploration and explanation process, the user can save and annotate interesting findings in the provenance bar. While tracking of the exploration process could also be automated, we decided to leave it to the user to directly filter important findings. The provenance bar, therefore, acts as a persistent cross-dashboard as well as cross-model digital blackboard and covers parts of the global monitoring and steering mechanisms (\autoref{subsec:global-monitoring-and-steering-mechanisms}), namely \emph{provenance tracking} and \emph{reporting \& trust building}.
\autoref{fig:provenance-use-case} shows the provenance track for an example model explanation and refinement process.

The reporting phase is the final phase of the frameworks workflow. Its goal is to offer a solution to common issues of missing justification, provenance tracking~\cite{Chen2018}, and reproducibility~\cite{Henderson2017}. In this prototype, we implement the reporting phase as an interactive arrangement of the provenance cards saved in the understanding, diagnosis, and refinement steps. This allows the user to see the feedback by the explainers he acted on, the decisions he made to refine, and, in the case of iterative loops through the workflow, the improved output of the repeated feedback from the explainers. By adding or modifying annotations, the user can document his thoughts and findings and, therefore, structure the process in a storytelling manner. This might be crucial if other people are involved in the model development or deployment process and, hence, justification or trust-building is a necessity. The reporting dashboard accordingly extends the functionality of the provenance bar by the global monitoring and steering mechanisms \emph{Comparative Analytics}, while further enhancing the \emph{storytelling} and \emph{justification} aspects.

\section{Evaluation}

In this section, we describe the methodology of our study, the feedback we received from the different target users, and the insights we extracted from the given feedback.

\subsection{User-Study}
To verify the intuitiveness of our workflow and the usability of the system, we conducted a qualitative user study with different types of target users. We use both a simple and a complex network trained on the MNIST dataset~\cite{Ciresan2011mnist},  simulating a real-life environment.
The goal of the study is to see where the system can be improved and whether all the necessary improvements are already covered in the framework, and, thus, only limitation of this specific system implementation.

\textbf{Methodology and Study Design -- }
Due to the variety of available interaction loops, we decided to conduct a pair analytics study~\cite{kaastra2014field}, enabling each participant to transfer their individual workflow to the system.
We performed nine approximately one-hour sessions in which a member of our team (henceforth referred to as visual analytics expert, VAE) worked with the target user (henceforth referred to as
model novice (MN, \inlinegraphics{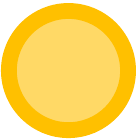}),
model user (MU, \inlinegraphics{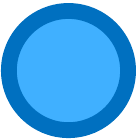}) and
model developer (MD, \inlinegraphics{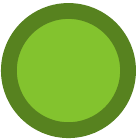})).
Each study started with a semi-structured interview regarding the user's previous experience with ML as well as their expectations on the framework and the system.
After gaining these unbiased insights,
the VAE gives a quick introduction to the system, available datasets, and the analysis tasks that the user should solve during the pair analytics session.
Then, the control of the system is handed over entirely to the participant.
They are asked to communicate their thoughts and actions by ``thinking aloud'' while conducting the predefined analysis tasks, taking as much time as they need.
The VAE can interrupt the session to clarify interaction possibilities, limitations, or to guide the user towards the next analysis task.
The last part of the study consists of another interview reflecting on the difference between the initial expectation and the experience during the pair analytics regarding the workflow, the system, and the performed analysis tasks.
All study sessions were audio-recorded and screen-captured.

\textbf{Participants -- }
We selected our participants from three different groups of target users. For the model novices (MN), we interviewed two Ph.D.\,students with a computer science background that had basic knowledge on ML but had never built NNs before. For the group of model users (MU), we interviewed two Ph.D.\,students with experience in the analysis of linguistic data but no prior experience with deep learning. For the model developers (MD), we interviewed five experts (two industry developers, three students) that were familiar with TensorFlow and TensorBoard. All participants had either finished or were currently pursuing a university degree. Only one of the participants was female, which could be explained by the low number of females in the domains we were recruiting from. 

\textbf{Tasks -- }
The participants were guided through the interaction along the tasks understand, diagnose, and refine, but were allowed to loop back.
In case the participants spent too much time on one task, the VAE would use unobtrusive questions to guide them to another task.

\subsection{User Feedback}
In the following, we describe the feedback received from participants during the three study phases (expectation, pair analytics, review).~\footnote{A graph of the complete feedback frequency by user group is given in Figure S1 (supplementary).
An overview of all the topics mentioned by each participant is given in Table S2 (supplementary).} The side-figures provided throughout this section summarize the frequent ($>1$) comments of participants, indicating their user group. In each section, we highlight the aspect that the side-figures address.
\newlength{\intextsepdefault}
\setlength{\intextsepdefault}{\intextsep}

\textbf{Expectations -- }
When we asked participants about the general utility and their expected \textbf{use-cases for XAI}, the most frequent answers
\setlength{\intextsep}{3pt}%
\begin{wrapfigure}[5]{r}{0pt}
  \includegraphics[scale=0.125]{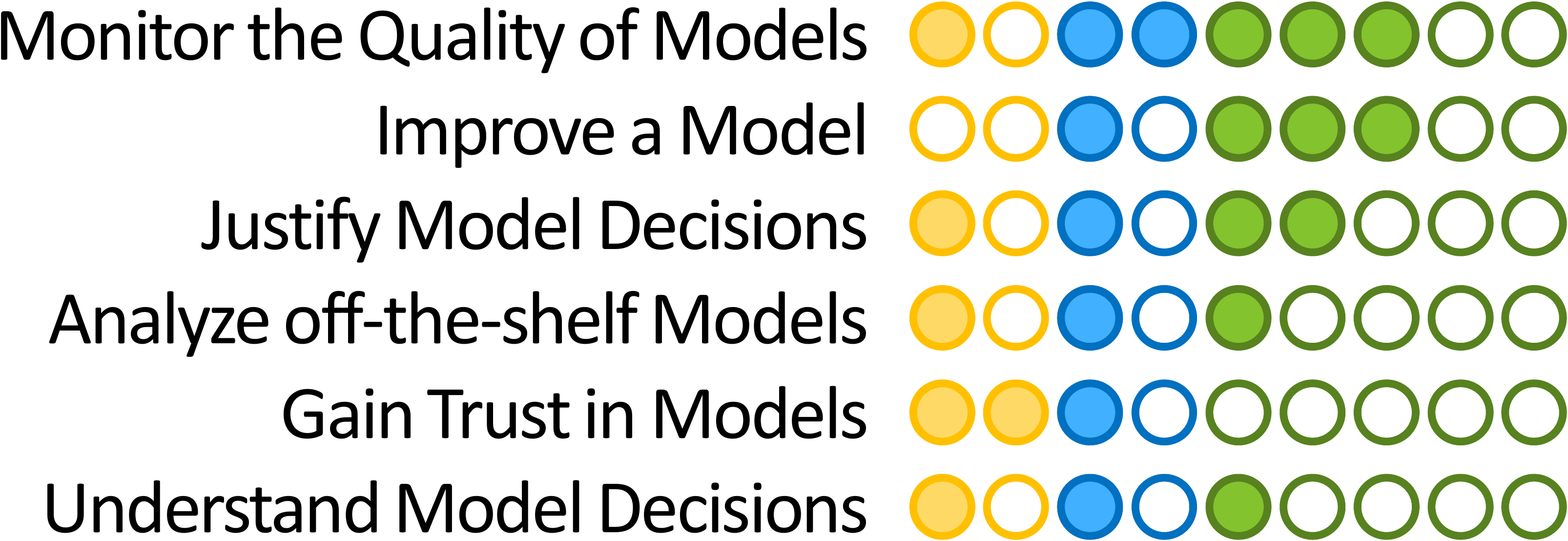}
\end{wrapfigure}
are in line with some of our frameworks global steering and monitoring mechanisms: model quality monitoring, search space exploration, reporting \& trust building, and knowledge generation. Additionally, the users from each user group wanted to verify pre-trained models with XAI methods, which is also supported by our XAI pipeline. Besides the most frequent suggestions, some users had extraordinary ideas, such as using XAI methods for marketing the model. Within our framework, this could be one manifestation of reporting \& trust building. A difference in ideas between user groups is that the MDs had more specific ideas (e.g., feature influence on decision) while the MNs and MUs mostly suggested high-level concepts (e.g., trust building).

When reviewing the suggested \textbf{framework}, most participants agreed with the tasks understanding, diagnosis, and refinement. How-
\setlength{\intextsep}{2pt}%
\begin{wrapfigure}[2]{r}{0pt}
  \includegraphics[scale=0.125]{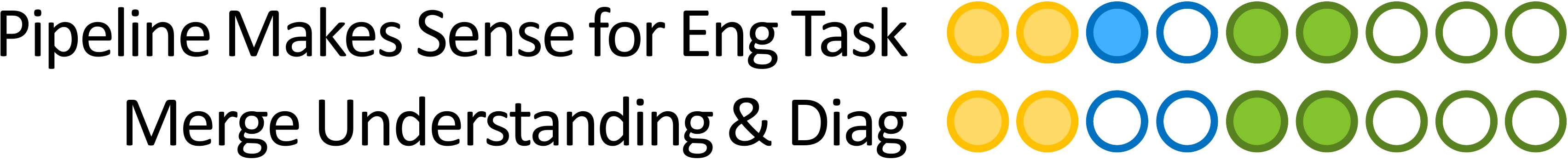}
\end{wrapfigure}
ever, many of them would also prefer to merge the understanding and diagnosis phase. This adaption is supported to some degree by our framework. Depending on the use-case, each task of the pipeline can be shortened, left out, or repeated. As we suggest in the description of our framework, the understanding task, for example, is often more important for MNs than for MUs or MDs. Besides the congruent feedback on the framework and pipeline, some individual ideas were presented, such as a separate model building task. While the model building is not a separate element in our pipeline, it can be simulated by starting with a minimal default model and continuous building-block-like refinements.

\textbf{Pair Analytics -- }
During the \textbf{understanding phase}, the users were first presented with a graph representation of their model. The graph
\setlength{\intextsep}{1.5pt}%
\begin{wrapfigure}[6]{r}{0pt}
  \includegraphics[scale=0.125]{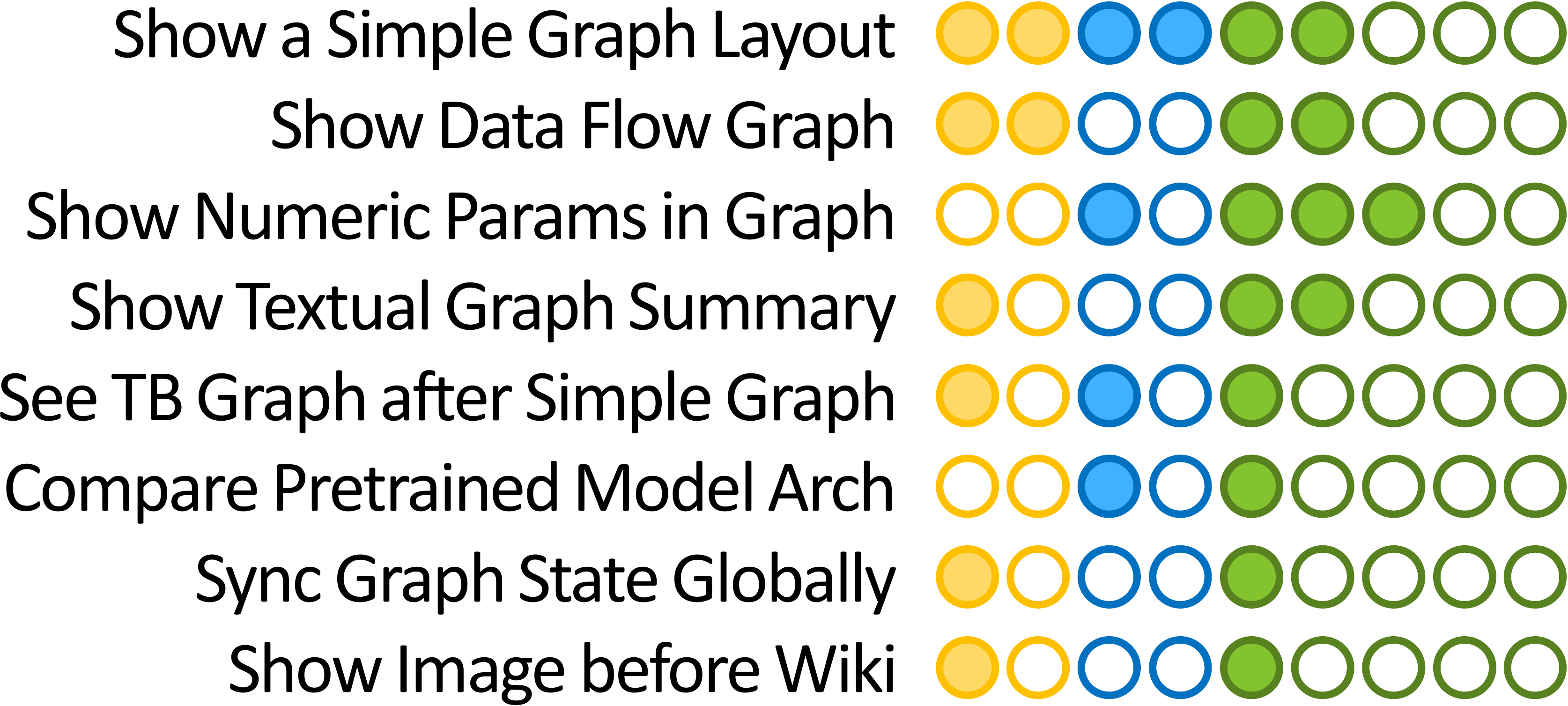}
\end{wrapfigure}
representation was criticized by many participants. This is one of the design decisions based on the integration into the TensorBoard environment and does not directly reflect on the framework. Furthermore, it suggests that, in parallel to the explainer toolbox, the graph represents another type of model content for which a toolbox of explanations (e.g., dataflow, classical layout, numeric parameters, text) should be offered. Some individual MDs even suggested showing the code as a representation and not needing the graph, if the model is self-built. Another difference that we see between different user groups is that the MDs focused more on details of the graph such as numeric parameters and code snippets than the MUs and MNs.

During the \textbf{diagnosis phase}, participants generally gave positive feedback. In addition to the provided explainers, users wanted to gain
\setlength{\intextsep}{2pt}%
\begin{wrapfigure}[4]{r}{0pt}
  \includegraphics[scale=0.125]{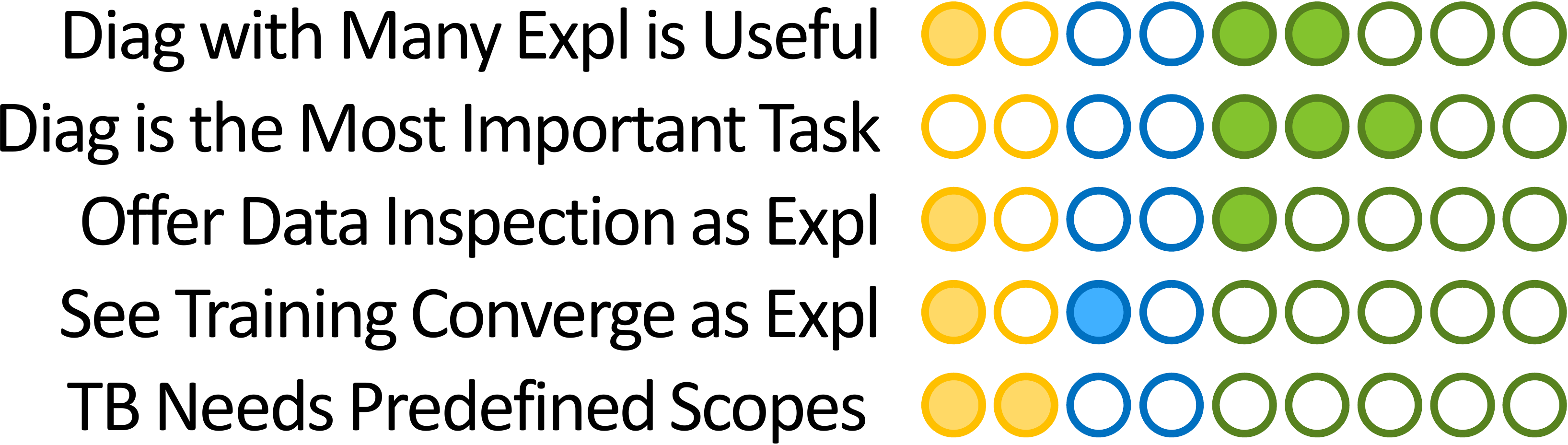}
\end{wrapfigure}
insight into the underlying data and other metrics of the model, such as convergence. Such features are instances of the different frameworks explainer types but have not been implemented in our system. For example, a model-agnostic explainer could review the dataset balance and show it to the user. Some further concerns were only affecting the implemented system, such as scalability and scopes for the calculation of an explainer output. A difference we see between user groups is that MDs value the diagnosis more than MUs and MNs.

Regarding the \textbf{toolbox set of} different low- and high-abstraction \textbf{explainers}, the most interesting insight was that trust in the explain-
\setlength{\intextsep}{1pt}%
\begin{wrapfigure}[6]{r}{0pt}
  \includegraphics[scale=0.125]{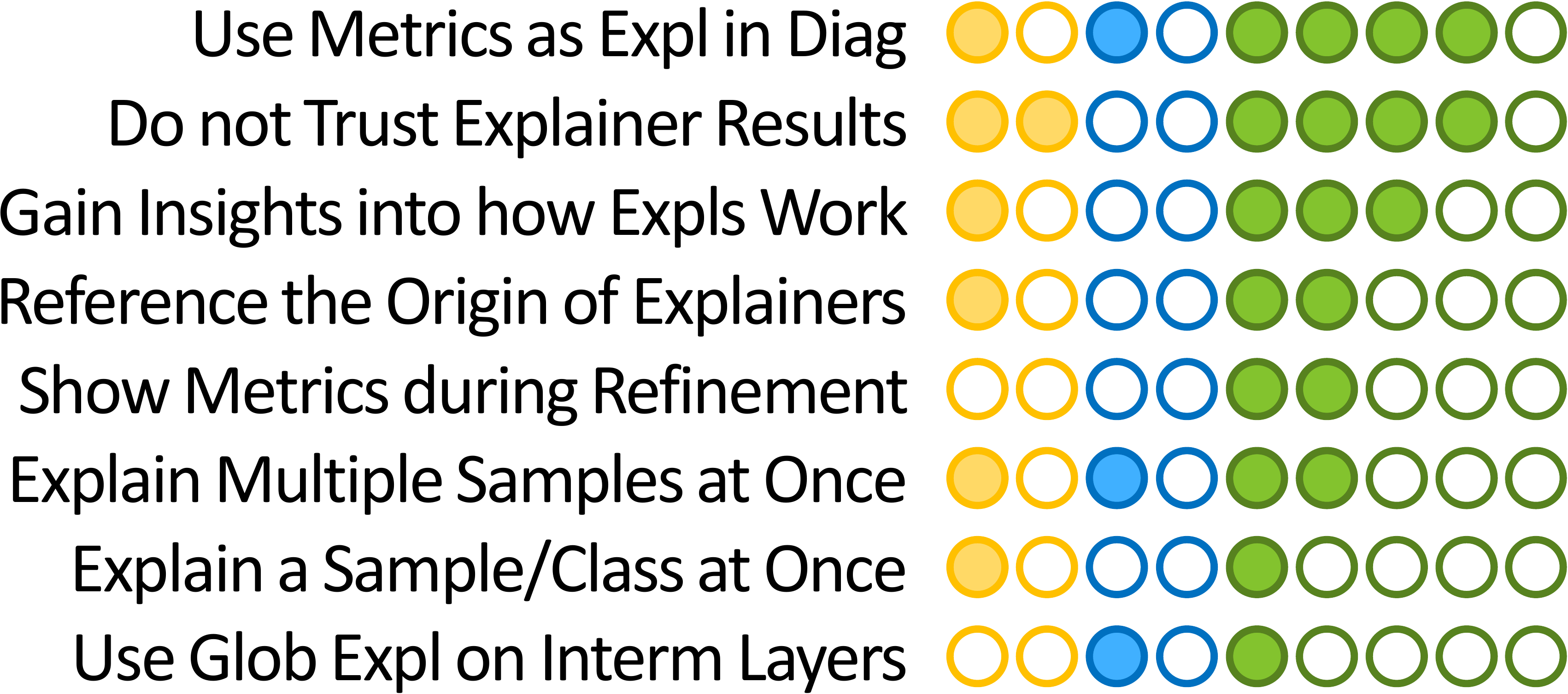}
\end{wrapfigure}
ers output is an issue for the users.
This aspect can, to some degree, be counteracted by giving more guidance within the system.
In the framework, this aspect is part of the targeted user guidance within the XAI strategies. In addition to the desired guidance, the participants wanted additional explainers, such as standard metrics and more low-abstraction explainers. Such additions can easily be made in future iterations of the system. A difference we see between user groups is that MUs were less concerned with trusting the explainer output than MDs and MNs.

During the \textbf{refinement phase}, the feedback and expectations were mixed. Some participants were very optimistic, suggested additional
\setlength{\intextsep}{3pt}%
\begin{wrapfigure}[5]{r}{0pt}
  \includegraphics[scale=0.125]{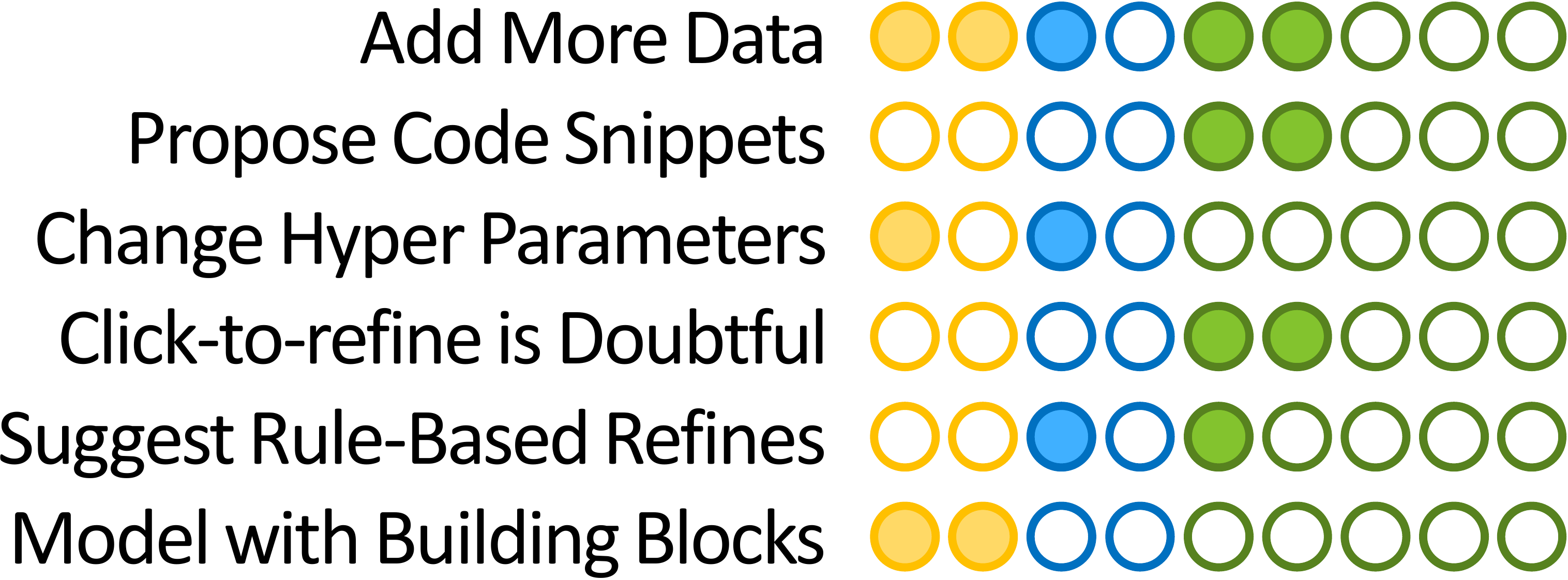}
\end{wrapfigure}
ways to interactively refine the model. All of the suggested refinements (e.g., adding data, changing parameters, switching architecture) are covered by the framework in the form of transition functions resulting from a single- or multi-model explainer. More doubtful participants did not criticize the utility of such a tool, but rather the possibility of offering this functionality with sufficient proficiency. In the future, the current systems should be extended with the suggested refinement methods and additional guidance to select the appropriate refinements.

Concerning the \textbf{model comparison}, many of the suggested improvements are in line with the global monitoring and steering
\setlength{\intextsep}{1.5pt}%
\begin{wrapfigure}[4]{r}{0pt}
  \includegraphics[scale=0.125]{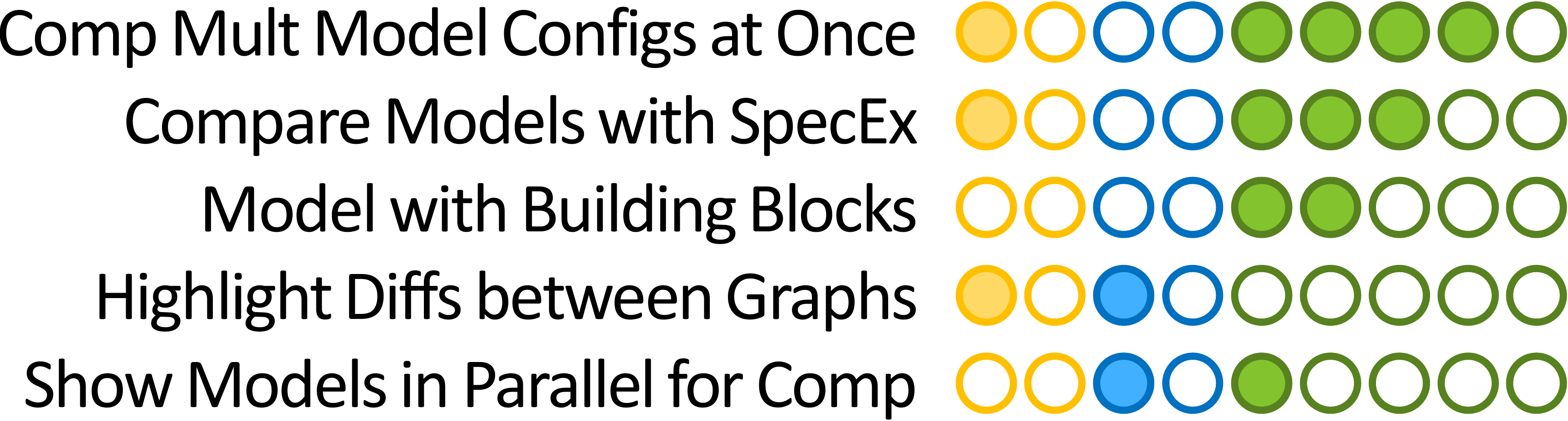}
\end{wrapfigure}
mechanisms of the framework, such as search space exploration, data shift scoring, and comparative analytics. They are an important part of future iterations of our system. Interesting individual ideas for future work were a building block system that could adapt the model architecture, the data and the features on the fly and compare different explainers/metrics on a selection of these models. This suggestion in line with a previous suggestion of having a separate model building task in the pipeline and can be simulated with an interactive interface for fast iterations of the refinement phase.

Concerning the overarching aspect of \textbf{provenance tracking and result reporting}, the feedback was very unified. All participants liked
\setlength{\intextsep}{2.5pt}%
\begin{wrapfigure}[9]{r}{0pt}
  \includegraphics[scale=0.125]{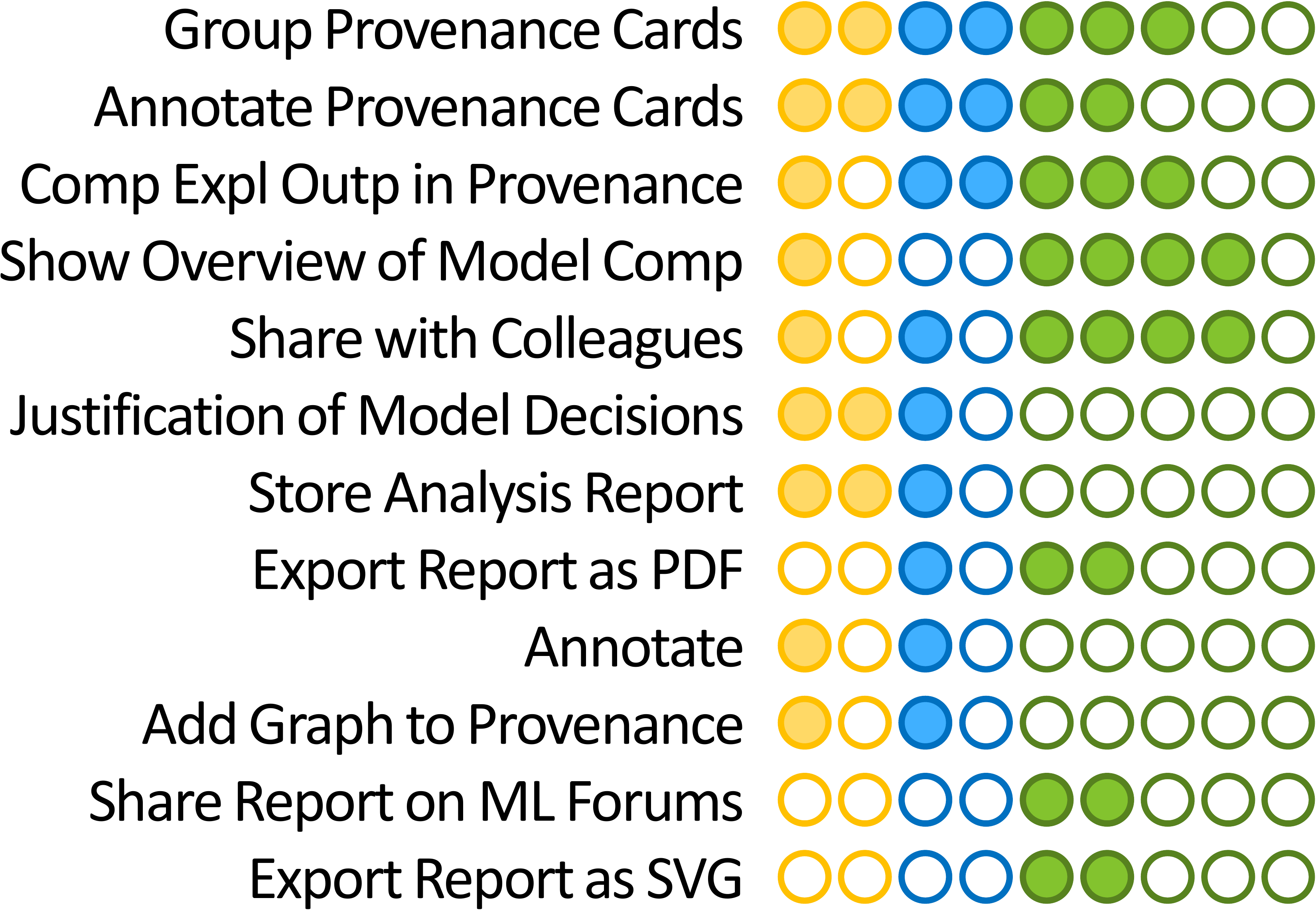}
\end{wrapfigure}
the feature and would use it to communicate their results.
The importance and acceptance of this feature further confirms the utility of the global monitoring and steering mechanism provenance tracking and reporting \& trust building. Within the system, users suggested many interactions, that should be added in future iterations, such as annotation tools, export formats, and layouts. Beyond the systems implementation, participants came up with several suggestions for utilizing this feature, such as different reports to colleagues or stakeholders.
A difference we see between user groups is that MUs and MNs target justification and MDs exchange between colleagues.
\setlength{\intextsep}{\intextsepdefault}

\textbf{Expectation Review -- }
Regarding the overall usability and value of the system for real use-cases, all participants gave very positive feedback.
Half of the participants considered the system too complex for beginners. Two participants stated that it would be okay for either model users or model novices. Only one participant explicitly stated it should only be used by developers. The experienced gap could be fixed by extending guiding mechanisms (see \autoref{subsec:future-work}), leaving a workflow which can capture the transition in user expertise while its distinct states can still be applied independently for specific tasks.
Common feature requests in this direction focus on additional user guidance in the form of tutorials, suggestions, checklists, information with more labels, and example models.

\subsection{Lessons Learned and Future Work}\label{subsec:future-work}

Overall, the feedback on the system design was positive. Additionally, the users had several ideas for complementary features, for which integration in the system is planned for future versions.
The most requested functionality was a simplified presentation of the model graph, with an option to switch to the more complex representation if required.
Another significant point was trust in the explanation methods themselves.
By including additional descriptions, links, and possible interpretations for every explainer, we tried to improve the confidence in the explanation.
This idea of \emph{meta-explanations} could be extended even further: instead of providing static descriptive content, dynamic visualizations could be rendered to explain the current explainer output, e.g., the surrogate-models architecture for LIME or a heatmap displaying the relevance in all layers for LRP.
Regarding the expert users, more advanced features were desired. Some users wanted to apply high-abstraction explanations on a subset of layers~\cite{Rauber16}.
Most commonly requested was an additional view to enable direct comparative analytics as well as the speculative execution of proposed refinements.
The refinement step was often rated to be the one with the highest potential, but it was also considered the most complex to implement.
Ideas to enhance its functionality included proposing code fragments, providing building blocks, or even scaling it up to a social platform were suggestions for model improvement could be shared among other developers. This could be extended by a way to restore saved exploration states or reproduce the process of exploration from other users.

By further extending user guidance, for example, by including AI driven recommendations, a broader range of user groups could be addressed.
The additional flexibility could make the tool suitable for educational purposes or more advanced analysis and refinement tasks.

While some of the user suggestions can be included in our existing system, e.g., a simplified graph view, others are not that easy to realize,
e.g., on-demand refinement.
While building upon TensorBoard saved us a significant amount of work for data logging, data loading, and developing the graph view, it also presented some limitations.
This concerns the user interface as well as export, storage, processing, and exchange of data.
The strict separation of tabs as \emph{plugins}, each with its distinct data backend, makes data sharing as well as mutual views (e.g., provenance bar) challenging to realize.
Furthermore, to close the IML loop of model development (TensorFlow) and model analysis (TensorBoard), those stages have to be combined.
While we were able to surpass some of these issues, e.g., by including an external backend or modifying the website template, a system which could provide a full IML and XAI workflow would require a more specialized architecture.

Finally, some users asked for additional, in particular low-abstraction, explanation methods. We deliberately designed the proposed framework, as well as the derived system, as a platform for the integration of already existing and entirely new explanation methods.

\section{Conclusion}

We presented explAIner, a framework for interactive and explainable machine learning, capturing the theoretical and practical state-of-the-art in the field. As a core concept of our XAI framework, we defined the XAI pipeline, which maps the XAI process to an iterative workflow of three stages: model understanding,  diagnosis, and refinement. By determining additional global monitoring and steering mechanisms, we extended the XAI pipeline by overarching tools and quality metrics. To show the practical relevance of our framework, we instantiated it in an actual system. Besides the three stages of the XAI pipeline, the implementation covers global monitoring and steering mechanisms by providing provenance tracking as well as an additional reporting step. To test the usability and usefulness of our tool, we performed a user study with nine participants, coming from different user groups. The users found our system to be intuitive and helpful and considered an integration in their daily workflow.

\acknowledgments{
This work has received funding from the European Union's Horizon 2020 research and innovation programme under grant agreements No 825041 and No 826494.
}

\clearpage

\bibliographystyle{abbrv}
\bibliography{ms}

\end{document}